\documentclass[letterpaper,twocolumn,10pt]{article}
\usepackage{usenix2019_v3}
\usepackage{comment}
\usepackage{times}

\usepackage{tikz}
\usepackage{amsmath}

\usepackage{filecontents}
\usepackage{subfloat}
\usepackage[subrefformat=parens,labelformat=parens]{subfig}
\usepackage{graphicx}
\usepackage{xurl}
\usepackage{multirow}
\usepackage{algorithm}
\usepackage[noend]{algpseudocode}
\usepackage{siunitx}

\newcommand{\parabf}[1]{\medskip\noindent\textbf{#1}}
\newcommand{\sysname}{PowerFlow}

\newcommand{\revise}[1]{\textcolor{black}{#1}}
\begin{document}
	%-------------------------------------------------------------------------------
	
	%don't want date printed
	\date{}
	
	\title{Energy-Efficient GPU Clusters Scheduling for Deep Learning}
	
	%for single author (just remove % characters)
		\author{
			{\rm Diandian Gu}\\
			Peking University
			\and
			{\rm Xintong Xie}\\
			Peking Univerisity
				\and
				{\rm Gang Huang}\\
			Peking Univerisity
				\and
				{\rm Xin Jin}\\
			Peking Univerisity
				\and
				{\rm Xuanzhe Liu}\\
			Peking Univerisity
			% copy the following lines to add more authors
			% \and
			% {\rm Name}\\
			%Name Institution
		} % end author
	
	\maketitle
	
	\begin{abstract}
Training deep neural networks (DNNs) is a major workload in datacenters today, resulting in a tremendously fast growth of energy consumption.
\revise{It is important to reduce the energy consumption while completing the DL training jobs early in data centers.} %\xzl{your paper does not focus on energy consumption, but the JCT improvement under energy budget. This statement is misleading.}
In this paper, we propose \sysname{}, a GPU clusters scheduler that reduces the average
Job Completion Time (JCT) under an energy budget. %\sysname{} finds the tradeoff between the average JCT and the total GPU energy consumption.
%reduces the average JCT of DL training jobs while achieving a goal of reducing the total GPU energy consumption.
We first present performance models for  DL training jobs to predict the throughput and energy consumption performance with different configurations. Based on the performance models, \sysname{} dynamically allocates GPUs and adjusts the GPU-level or job-level configurations of DL training jobs. \sysname{} applies network packing and buddy allocation to job placement, thus avoiding extra energy consumed by cluster fragmentations. Evaluation results show that under the same energy consumption, \sysname{} improves the average JCT by 1.57 - 3.39$\times$ at most, compared to competitive baselines.

%Not more than 200 words, if possible, and preferably closer to 150.
\end{abstract}
	\section{Introduction}

Deep learning (DL) powers many applications and services
we use every day. Therefore, training deep neural networks (DNNs) has become an important workload in datacenters. To increase the accuracy of DNN models, a substantial volume of training data is required; and the DNN model architectures become more and more complex (e.g., BERT large~\cite{DevlinCLT19} with 340 million parameters and GPT-3~\cite{gpt3} with 175 billion parameters). It is reported that the computation demand of DL training grows at a speed of 35$\times$ every 18 months~\cite{openai}.

However, this growing demand for computation leads to greater energy demand and carbon emissions.
For example, the estimated CO$_2$ emissions of training a Transformer (large) model are 5 times the life cycle CO$_2$ emissions of the car~\cite{transformerCarbon}. Many service providers have made their goals to reduce their carbon footprints~\cite{microsoft, google, facebook}. As an energy-consuming workload, reducing the total energy consumption of DL jobs is of vital importance to achieving such goals. Meanwhile, for the service providers' revenue, reducing the job completion time (JCT) of DL training jobs as much as possible is also substantial. It is beneficial for service providers to speed up the DL training jobs under a given energy budget, \revise{as this is not only energy-saving and environmentally friendly, but also leaves resources to serve more jobs.}%\xzl{Why intuitive? Under a given budget, reducing the JCT means that the service providers can allocate more resources for other newly incoming jobs? The benefits seem to be a bit not that straightforward. }

As far as we know, the current practice for scheduling
DL jobs is neither energy-aware nor energy-efficient. Most schedulers for DL jobs ignore the GPU-level configurations such as \textit{GPU core frequency}, which has a significant impact on the throughput and energy consumption of DL jobs. Many schedulers execute DL jobs with user-defined job-level hyperparameters, which might not be energy-efficient. Also, existing GPU schedulers ignore the energy-consumption characteristics of different DL jobs, leading to a waste of energy in the cluster. 
Several efforts have studied the impact of GPU Dynamic Frequency and Voltage Scaling (DVFS) and power configuration on the energy consumption and the throughput of DNN training~\cite{BharadwajDEOK21, KomodaHNMN13, TangWWC19}. But directly applying DVFS or power configuration lacks the overall perspective from the whole GPU cluster and cannot leverage elasticity to further optimize the energy consumption and average JCT.
\revise{There are virtual machine (VM) schedulers to reduce energy consumption in traditional clusters~\cite{WangLL21, DingQLW15}. These methods require fine-grained resource allocation. However, the common practice for DL training jobs dedicates the whole GPU to a single job. The fine-grained methods do not apply to this common practice.}

%\xzl{This para does read more relevant to reducing the energy, but this not the focus of this paper, right? There is a huge logical gap here. I guess that you may want to state: 1. energy consumption is a big issue; 2. in some scenarios, the service provider may have a given energy budget for a set of need-to-process jobs; 3. an intuitive requirements of service provider may want to maximize the resource consumption and speed up the JCT under a given energy budget. Although it seems to be reasonable at first sight, the story  is not that apparently strong. What is the key benefit for the service provider? We should provide more convincing benefits to enhance our motivation.}

%\xzl{Also, even though we can clarify the motivation is reasonable, what is the key challenge? You emphasize on the GPU energy consumption. However, traditional data centers or clusters that rely on CPU, may have similar problems of energy saving? What is the key difference? }

In this paper, we present \sysname{}, an energy-aware scheduler for DL training jobs in GPU clusters. Under a given energy budget (i.e., the total consumed energy does not exceed a certain goal in a period of time), 
\sysname{} tries to tradeoff between the average JCT and the total energy consumption of the submitted DL training jobs.
%With a goal of consuming a specific amount of energy in the cluster for a time slot (i.e., energy budget), we optimize for the average Job Completion Time (JCT) of the cluster.
Compared with existing solutions, \sysname{} holds the view of the whole cluster and adjusts both GPU-level and job-level configurations dynamically to make the whole cluster energy-efficient.

Designing such a solution can be challenging. First, it is difficult to know how throughput and energy consumption change with different configurations. Existing studies only summarize coarse-grained features of DL jobs' performance: 
\revise{some of them only consider a fixed configuration for a DL job~\cite{abs-2007-03051}, while others do not consider the GPU-level configurations such as GPU frequency~\cite{QiaoCSNH0GX21}.
Therefore, existing studies are not sufficient and accurate to predict the performance of DL training jobs whose configurations change dynamically.}
To accelerate DL jobs under a specified energy budget, we need to know how the throughput and energy consumption change with different configurations.
Second, there is a tradeoff between the energy consumption and the JCT of a DL job.
In a cluster with limited GPU devices, it is difficult to decide whether a DL job should use less energy to achieve the energy budget goal or use more energy for a shorter JCT.

To address these challenges, we model how a DL job's performance (i.e., throughput and energy consumption) changes with different GPU-level and job-level configurations. Then, we propose \textit{energy efficiency} to capture how much throughput improvement can be brought by each unit of energy consumption. Based on the performance model and the energy efficiency metric, we design a greedy algorithm to prioritize resource allocation and GPU frequency raise for the most energy-efficient job without violating a \textit{power limit}. \sysname{} applies network packing and buddy allocation to job placement, avoiding extra energy consumption of cluster fragmentations.

In summary, we make the following contributions:
\begin{itemize}
	\item We show that a model of a DL job’s throughput and energy consumption per iteration can be learned by observing its step time and energy consumption during training and be used for predicting the performance given different configurations.
	\item We propose a formulation of energy efficiency for DL jobs in a cluster, which is a measure of training performance that takes into account both training throughput and energy consumption.
	\item We design and implement a scheduling algorithm that uses such performance models and the energy efficiency metric to schedule each submitted DL job.
	\item We evaluate \sysname{} with simulation experiments using a 1901-job trace. Experiment results show that  \sysname{} improves the average JCT by 1.57 - 3.39$\times$ at most, compared to competitive baselines. This indicates that \sysname{} is energy efficient.
\end{itemize}

	\section{Background}
\subsection{DNN Training}
%How to train DNN models
A DL training job trains a DNN model with a dataset. The job includes many iterations, and each uses a batch of samples from the dataset to train the model with a forward propagation pass and a backward propagation pass. The batch size is a hyperparameter set by DL training developers, which affects not only the model accuracy but also the training throughput and energy consumption. Usually, the DNN models are trained on powerful accelerators such as GPUs.

As DNN models are widely used in various applications and services~\cite{BerardPSB16, ChenSKX15}, training DNN models has become an important workload in datacenters. 
To improve the accuracy of DNN models, the training datasets and the sizes of models are growing larger and larger. Therefore, it requires more and more computation resources for DL training.
Since 2012, the amount of computation used in the largest AI training runs has been increasing exponentially with a 3.4-month doubling time~\cite{StrubellGM20}.  It is time-consuming to train such large DNN models, so distributed training is widely used to speed up the training process. In the data parallelism~\cite{KrizhevskySH12, li2014scaling, li2014communication, paszke2019pytorch} strategy of distributed training, the batch size of each worker is called local batch size, and the sum of the batch size of all workers is called global batch size. There are also other strategies, such as model parallelism~\cite{DeanCMCDLMRSTYN12}, hybrid parallelism~\cite{JiaZA19}, and pipeline parallelism ~\cite{HuangCBFCCLNLWC19}. In this paper, we leverage data parallelism with elastic training to adjust the throughput and energy consumption of DL training jobs.

The growth of large training datasets and large DNN models leads to enormous energy consumption.
It is reported that most of the energy consumption in data centers is in servers~\cite{masanet2020recalibrating}. Recent benchmarks show that among hardware devices such as GPU, CPU, and DRAM, GPUs are responsible for about 70\% of the total energy consumption during DNN training~\cite{LiCBZ16, DodgePCOSSLSDB22, HodakGD19}. Therefore, we focus on the energy consumed by GPU devices in this work. 

\subsection{Limitations of Existing Solutions}
%Usually, DL training developers wish to finish their jobs as soon as possible. The period of time from submitting a DL training job to the job's completion is measured in terms of JCT. 
 %\xzl{A single job can also be scheduled across clusters. Do you mean that Zeus
%	 lacks the consideration of running multiple jobs?}
Some schedulers for VM have been proposed to reduce energy consumption in traditional clusters~\cite{WangLL21, DingQLW15}. These schedulers require fine-grained resource allocation such as allocating the CPU cores to different VMs. However, these efforts cannot be directly applied to a GPU cluster. This is because GPU sharing is still undeveloped: it brings many problems~\cite{LimAXKJ21} such as performance degrade, memory corruption, error propagation, etc.
The common practice for DL training is to dedicate a GPU to a single job.

%GPU sharing degrade the DL jobs' performance and the jobs on the same GPU device might encounter memory corruption~\cite{LimAXKJ21}.
%Therefore, the common practice for DL training jobs does not share GPUs.

Considering the characteristics of GPU devices and DL jobs, schedulers for DL training jobs in GPU clusters are proposed ~\cite{XiaoBRSKHPPZZYZ18, GuCSZJQLG19, HwangKKSP21, QiaoCSNH0GX21}.
These schedulers often optimize for the goal of minimizing average JCT by adjusting the execution order of training jobs or allocating different numbers of GPUs to each job. However, none of these schedulers consider the total energy consumption of DL training jobs. Zeus~\cite{abs-2208-06102} navigates the tradeoff between energy consumption and performance optimization for a single DL training job, but \revise{lacks the consideration of running multiple jobs from the perspective of the whole GPU cluster.}
	\section{Motivation}

\subsection{Opportunities}
There are opportunities for reducing the average JCT under an energy budget from three aspects: GPU level, job level, and cluster level.

From the perspective of a single GPU, the default  GPU core frequency is usually the largest supported frequency, which is not energy-efficient~\cite{abs-2208-06102}. 
The frequency of a GPU has a great influence on the power of the GPU, thus influencing the training throughput and energy consumption of a training job.
%The GPUs are usually set at the maximum frequency by default. 
DVFS tunes the frequency and voltage of hardware devices and can be adopted by laptop computers, servers, and mobile devices to conserve energy~\cite{Mei0017}.
The CPU DVFS technology is well-developed and has been adopted in both personal computing devices and large-scale clusters~\cite{dvfs_cpu}. Despite the maturity of CPU DVFS, the study of GPU DVFS is still at an early stage.
Therefore, the GPU cluster scheduler can set the GPUs at a more energy-efficient frequency than the default settings according to the energy budget of the whole cluster. 

From the perspective of a DL training job, the job-level hyperparameters may not be energy-efficient. 
DL training developers are likely to be familiar with DL algorithms (e.g., designing the architecture of DNN models, setting the global batch size, etc.) but have limited expertise in systems (e.g., choosing a local batch size that fits in the GPU memory, setting the number of GPUs for training, etc.).
Compared to the hyperparameters set by developers, there might be another set of hyperparameters that achieves a smaller JCT or less energy consumption. 
For example, for a DL training job that trains the VGG16 model with a global batch size of 64 on NVIDIA A100 GPUs, it takes \SI{0.112}{s} and \SI{39}{J} to train one iteration on two GPUs, but it only takes \SI{0.114}{s} and \SI{27}{J} on one GPU\footnote{We measured the GPU energy consumption with NVML~\cite{nvml}}. Compared to the two-GPU setting, the one-GPU setting consumes 31\% less energy with a step time loss of only 2\%. Therefore, we can leave the configuration of system-related configurations to the scheduler (i.e., adjust the number of GPUs dynamically and adjust the local batch size accordingly).

From the perspective of a GPU cluster, 
existing solutions do not consider the energy impact of their scheduling decisions.
On one hand, for schedulers that do not leverage the elasticity of DL jobs (e.g.,  the number of GPUs does not change during job execution)~\cite{GuCSZJQLG19, XiaoBRSKHPPZZYZ18}, there might be fragmentations in the cluster, and the idle GPUs consume non-negligible energy. 
On the other hand, the schedulers with elastic resource allocation~\cite{QiaoCSNH0GX21,HwangKKSP21} ignore the different energy consumption patterns of different jobs, thus resulting in energy inefficiency. An energy-efficient scheduler should try to avoid fragmentations in the cluster while being aware of the different characteristics of different jobs.
\begin{figure}
	\centering
	\subfloat[GPU allocation with Tiresias over time.]{
		\begin{minipage}[b]{0.55\linewidth}
			\includegraphics[width=1\linewidth]{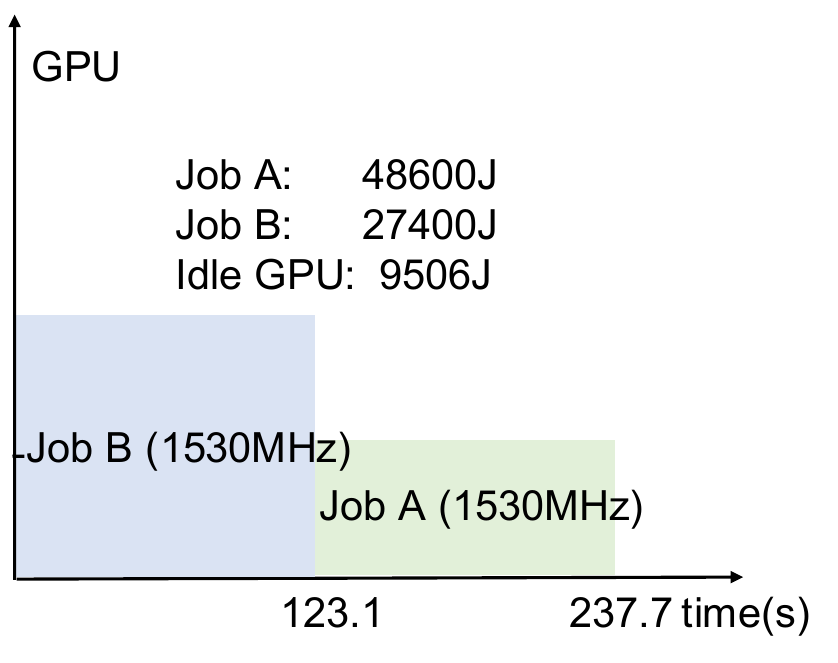}
		\end{minipage}
	}
	\subfloat[Optimal schedule.]{
		\begin{minipage}[b]{0.45\linewidth}
			\includegraphics[width=1\linewidth]{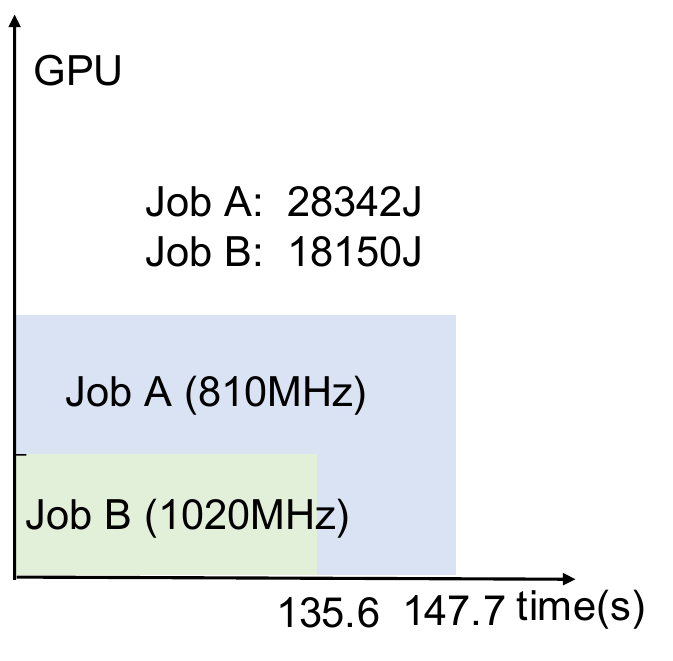}
		\end{minipage}
	}
	%\vspace{-0.1in}
	\caption{Motivating example to show that existing DL schedulers are not energy-efficient.}
	%\vspace{-0.05in}
	\label{fig:motivating_example}
\end{figure}

Combining the above three opportunities, there is still room for improvement in reducing JCT with an energy consumption budget.
Consider two DL training jobs in a cluster with two A100 GPUs.  Job A trains a VGG16 model with global batch size 64. The developer sets the number of GPUs to 1. 
Job B trains a GPT2 model with global batch size 32. The developer sets the number of GPUs to 2.
Both jobs are submitted at the same time and are set to train for 1000 iterations. The scheduling result with Tiresias~\cite{GuCSZJQLG19} scheduler is shown in Figure~\ref{fig:motivating_example}(a). The average JCT is \SI{180.4}{s} and the total energy consumption (including the energy consumption of the idle GPU) is \SI{85546}{J}. Figure Figure~\ref{fig:motivating_example}(b) shows a scheduling result with smaller JCT and less energy consumption: the JCT is reduced by 21\% and the energy consumption is reduced by 46\%. The improvement is achieved by adjusting the GPU frequency and the number of GPUs dynamically, and the specific configurations depend on the performance characteristics of the jobs.
	\section{Performance Modeling}
\begin{figure*}
	\centering
	\subfloat[ResNet18]{
		\begin{minipage}[b]{0.19\linewidth}
			\includegraphics[width=1\linewidth]{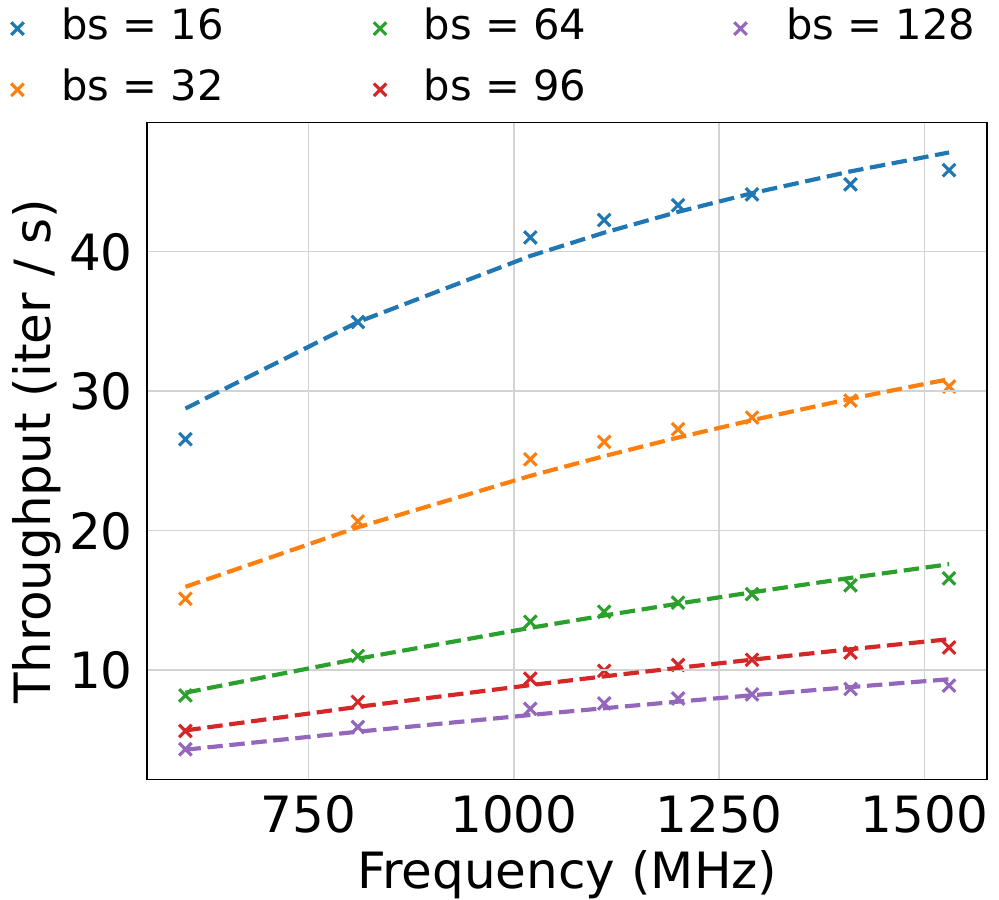} \\
			\includegraphics[width=1\linewidth]{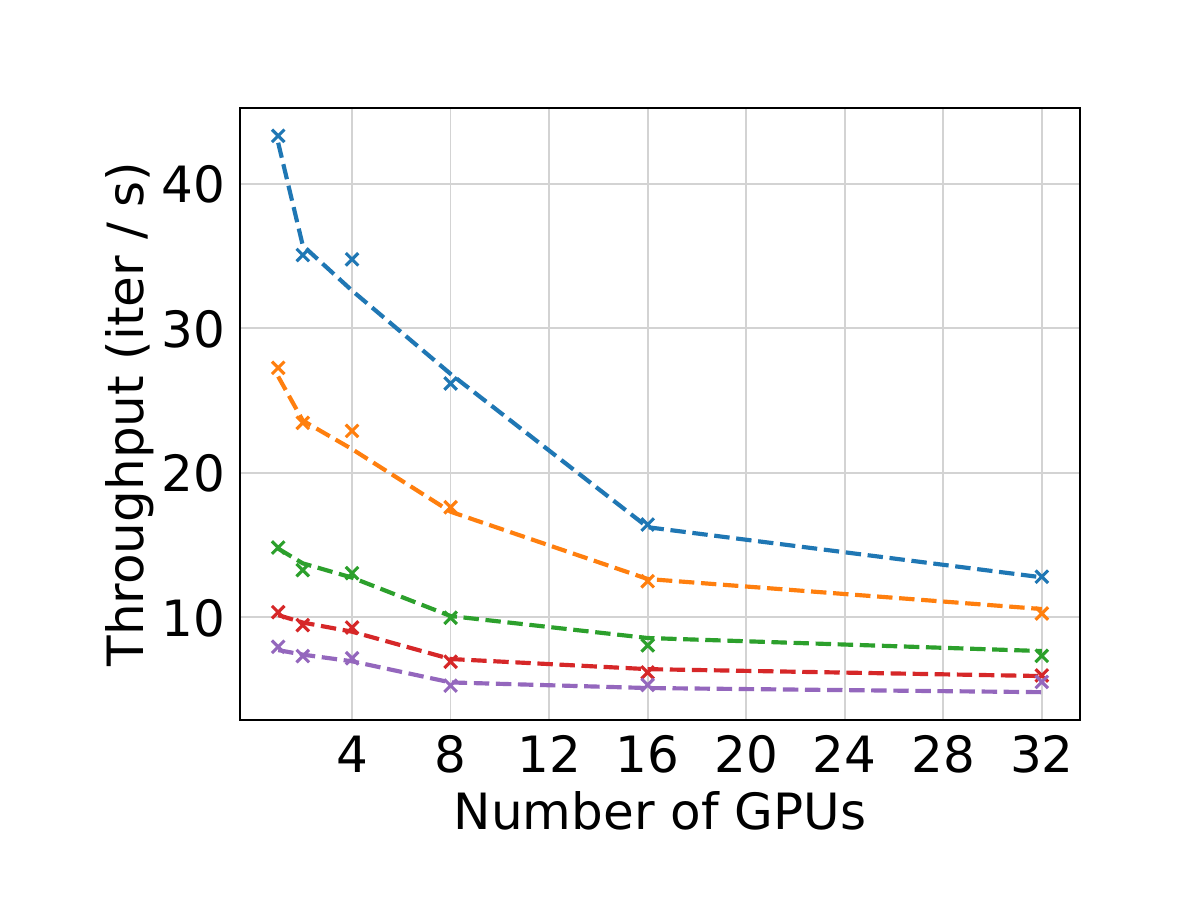}
		\end{minipage} 
	}
	\subfloat[VGG16]{
		\begin{minipage}[b]{0.195\linewidth}
			\includegraphics[width=0.98\linewidth]{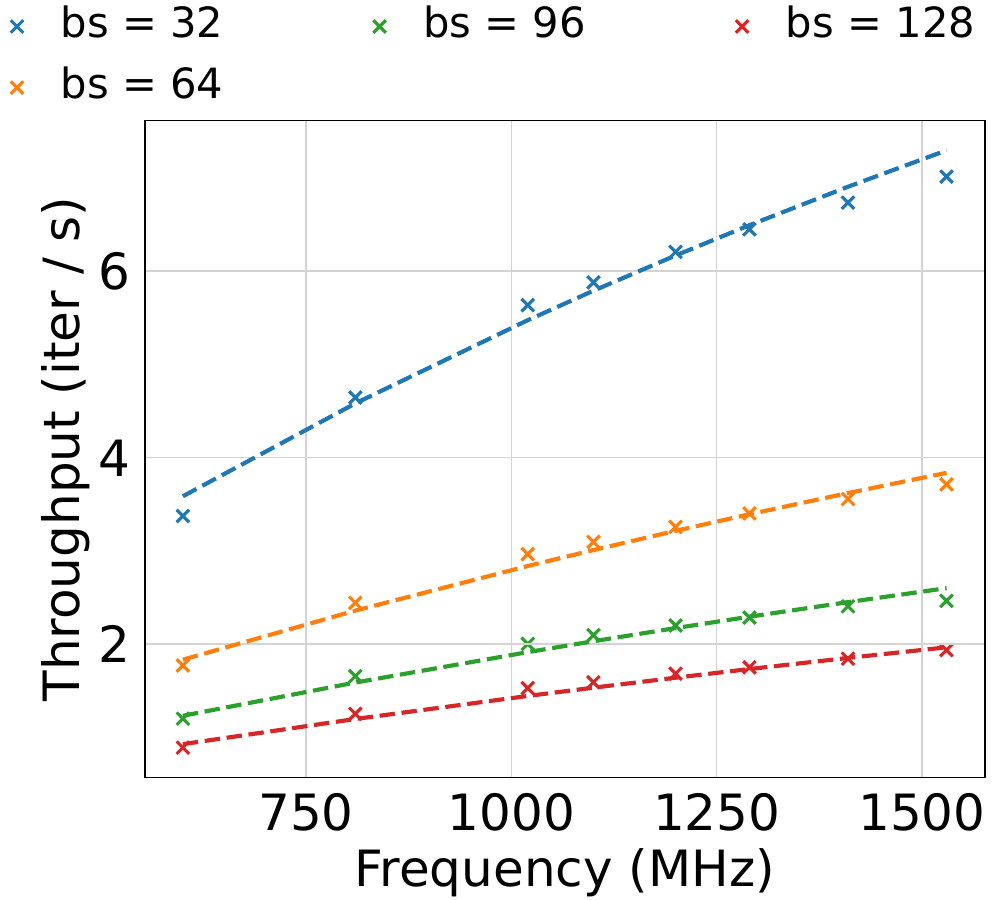} \\
			\includegraphics[width=0.97\linewidth]{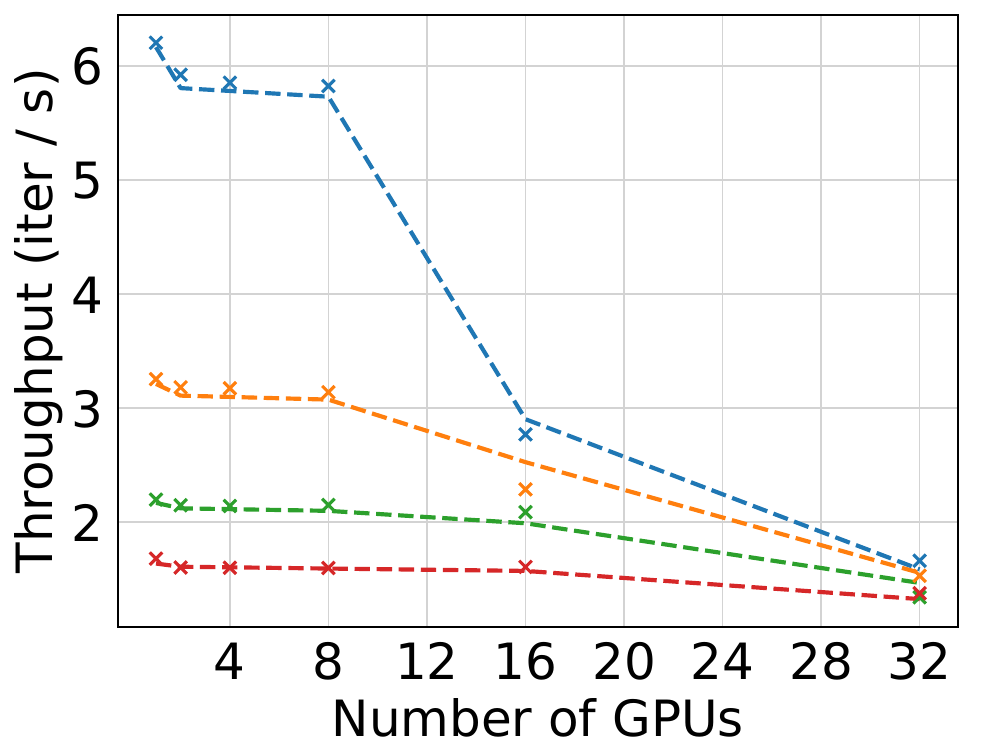}
		\end{minipage} 
	}
	\subfloat[Inception V3]{
		\begin{minipage}[b]{0.19\linewidth}
			\includegraphics[width=1.02\linewidth]{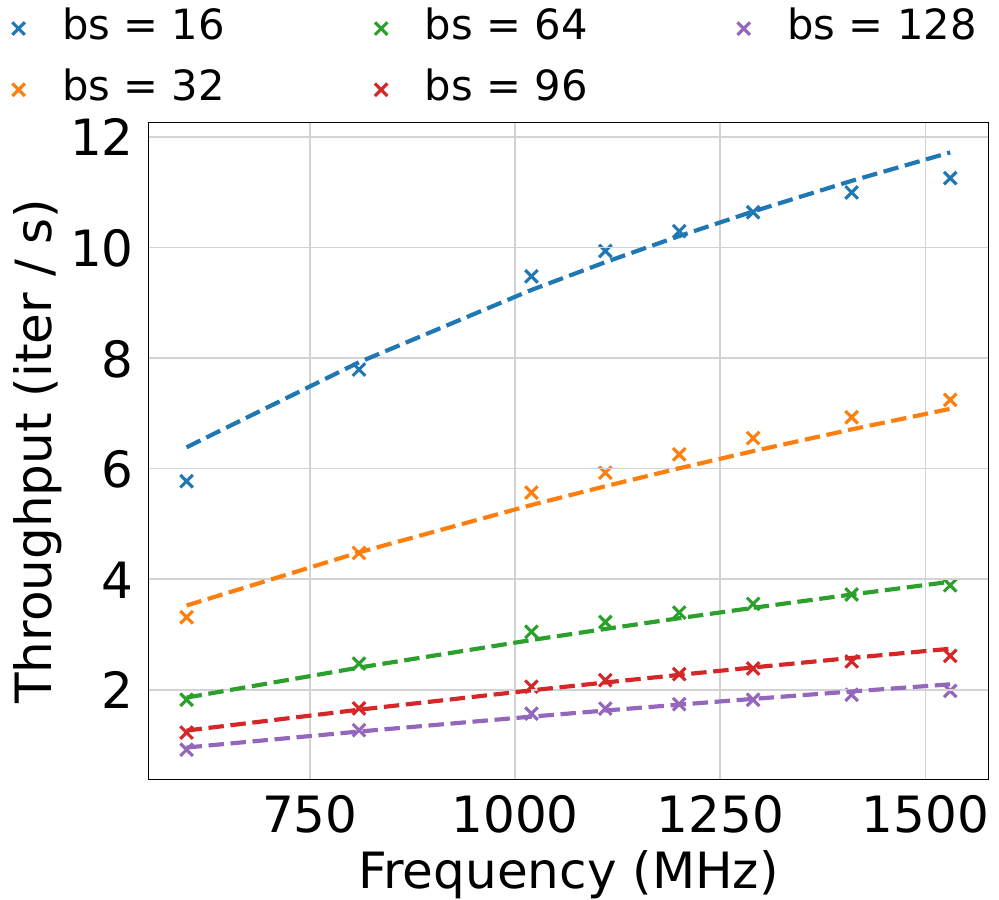} \\
			\includegraphics[width=1\linewidth]{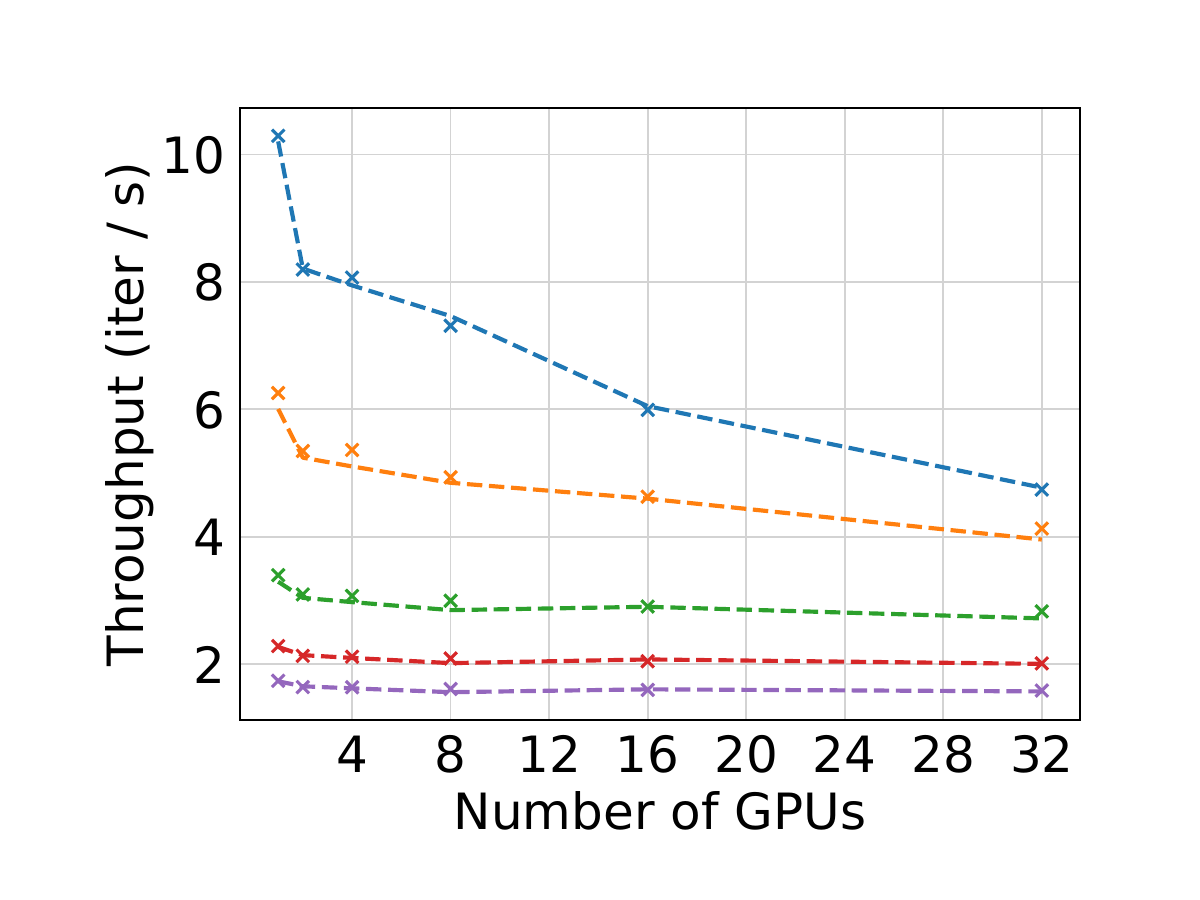}
		\end{minipage} 
	}
	\subfloat[GPT2]{
		\begin{minipage}[b]{0.19\linewidth}
			\includegraphics[width=1\linewidth]{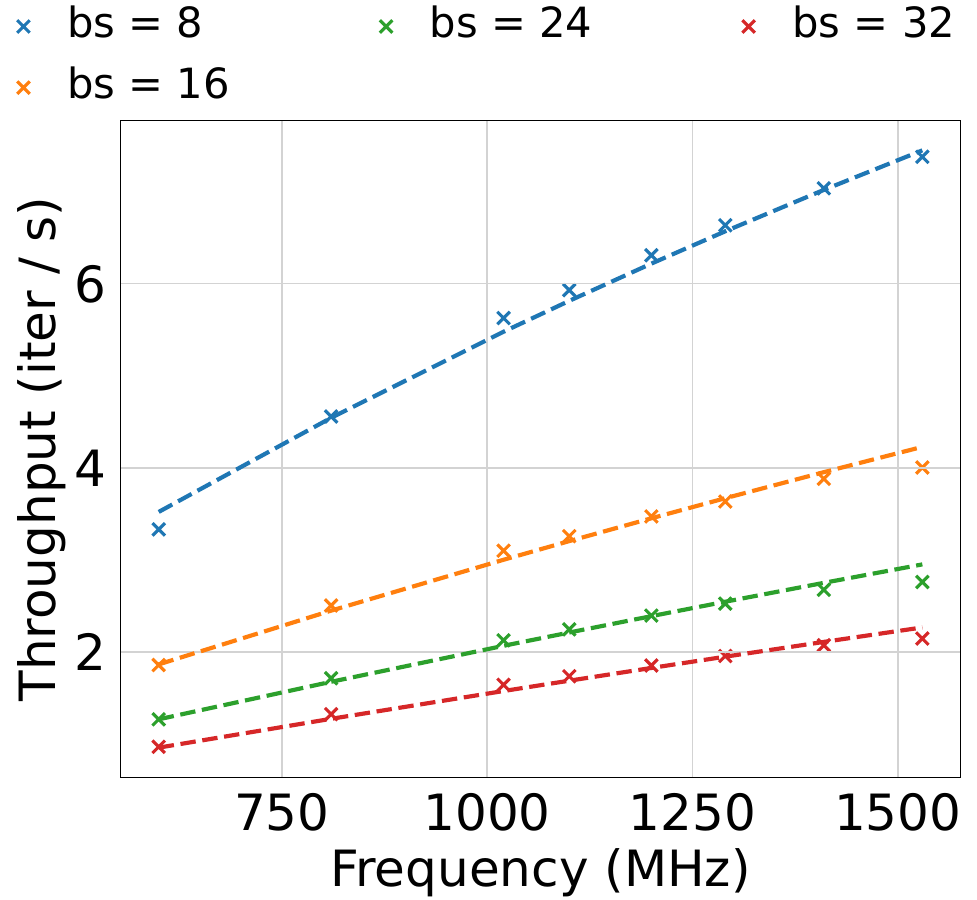} \\
			\includegraphics[width=0.95\linewidth]{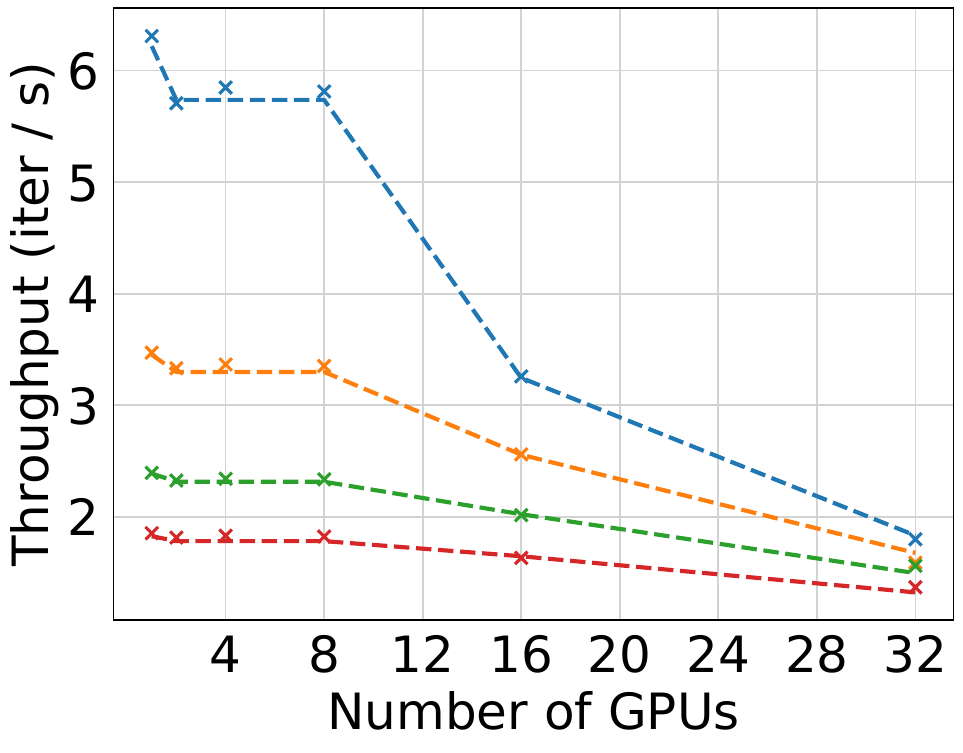}
		\end{minipage} 
	}
	\subfloat[Deep Speech 2]{
		\begin{minipage}[b]{0.19\linewidth}
			\includegraphics[width=1\linewidth]{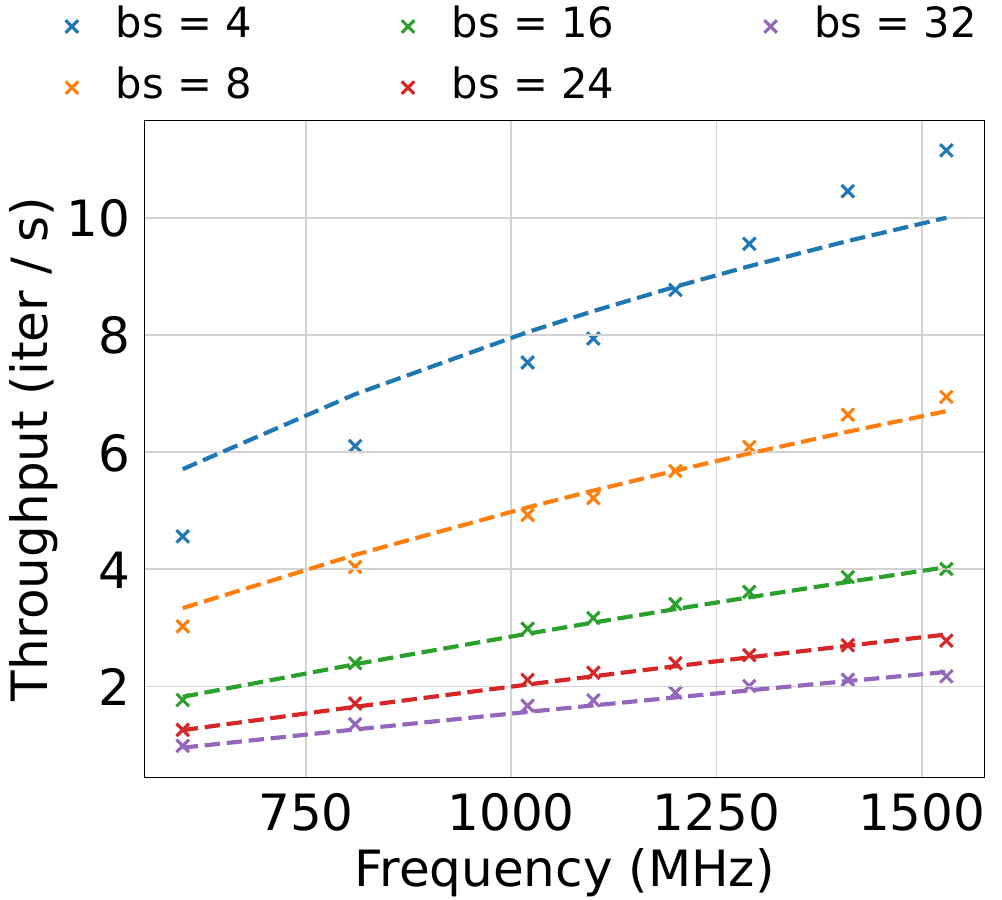} \\
			\includegraphics[width=1\linewidth]{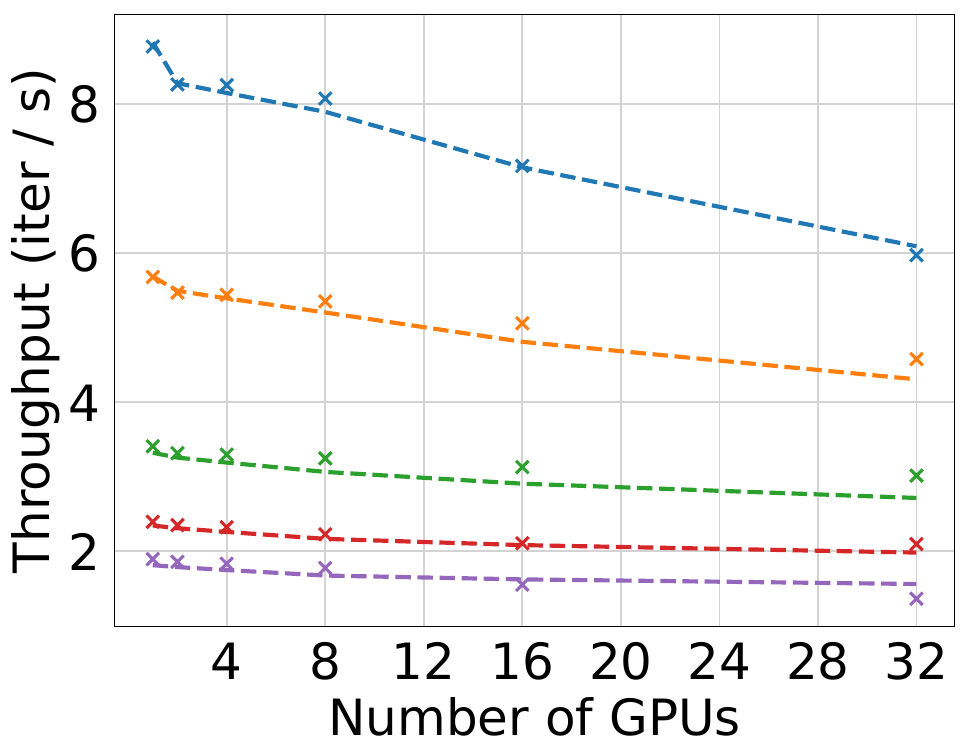}
		\end{minipage} 
	}
	%\vspace{-0.1in}
	\caption{Throughput of different DNN models with different configurations. The scattered data points are the throughput values measured on real NVIDIA V100 GPUs and the lines show the values predicted by the models. The figures on the top show how throughput changes with GPU frequency on one GPU. The figures at the bottom show how throughput changes with the number of GPUs with \SI{1200}{MHz} frequency. Note that the local batch size is kept the same and the global batch size scales linearly with the number of GPUs.}
	%\vspace{-0.2in}
	\label{fig:step_time_model}
\end{figure*}

In this section, we illustrate that the performance (including the training throughput and energy consumption per iteration) of DL training jobs can be measured during training and used as predictive models.  \sysname{} leverages these performance models to tradeoff between JCT and energy consumption under an energy budget. \sysname{} maintains the same algorithm-related hyperparameters  (i.e., global batch size, etc.) and adjusts the other system-related configurations. We focus on three system-related configurations of DL jobs:
\begin{itemize}
	\item $n$: the number of GPUs allocated for the DL training job.
	\item $bs$: the local (i.e., per-GPU) batch size, which is defined as $bs = BS / n$, where $BS$ is the global batch size specified by the DL developer.
	\item $f$: the core frequency of GPUs. 
\end{itemize}

\subsection{Modeling Throughput}
\label{sec:model:tpt}
First, we model and predict the throughputs of training a DNN model under different configurations. The throughput is defined as the number of iterations that can be executed in each unit of time:
\begin{align}
	\label{eq:tpt}{tpt} = 1 / T_{iter},
\end{align}
where $T_{iter}$ is the step time (i.e., training time per iteration) for a DL training job. 
DL training requires multiple types of resources such as GPU, CPU, network, etc~\cite{ZhaoLPZLJ22}. 
Therefore, we separately model  $T_{IO}$ for the time in each iteration spent reading the dataset from disk, $T_{grad}$ for the time in each iteration computing gradients locally, and $T_{sync}$ for the time in each iteration synchronizing model parameters between different GPU devices.

\parabf{Modeling $T_{IO}$.} 
%Storage IO from local or remote storage for reading training data into workers is required in DL training. 
In DL training, reading training data into workers requires storage IO from local or remote storage.
It is common to prefetch the training samples of the next batch while training the current batch (i.e., pipeline storage IO and GPU computation). We model $T_{IO}$ as a linear function of the local batch size. GPUs co-located on the same node might compete for the IO bandwidth of the node. Therefore, we also include a linear factor to model $T_{IO}$ with the number of allocated GPUs on each node. Thus, $T_{IO}$ is modeled as:
\begin{align}
	T_{IO} = \alpha_{IO} + \beta_{IO} * bs * r,
\end{align}
where $r$ is the number of allocated GPUs on each node and $\alpha_{IO}$ and $\beta_{IO}$ are fittable parameters. Note that we assume the number of GPUs on every node is the same.

\parabf{Modeling $T_{grad}$.}
The time spent on computing local gradients with forward and backward propagation is relevant to the GPU core frequency $f$ and the local batch size $bs$.
On one hand, by definition, the amount of computation executed in a certain time slot scales linearly with the device frequency. Therefore, we also model $T_{grad}$ as inversely proportional to $f$.
On the other hand, $T_{grad}$ scales linearly with $bs$.
The final  $T_{grad}$ is modeled as:
\begin{align}
	T_{grad} = \alpha_{grad} + (\beta_{grad} + \kappa_{grad} / f) * bs,
\end{align}
where $\alpha_{grad}$, $\beta_{grad}$, and $\kappa_{grad}$ are fittable parameters.

\parabf{Modeling $T_{sync}$.}
Following previous work~\cite{QiaoCSNH0GX21}, we separately model $T_{sync}$ when the job is allocated on a single GPU (no synchronization is required), on multiple GPUs of the same node (synchronize with PCIe or NVLink), and on multiple nodes (synchronize with Ethernet or InfiniBand). Similar to  $T_{grad}$, we estimate the communication time between GPUs as inversely proportional to $f$:
\begin{align} 
	\footnotesize
	T_{sync}=\left\{
\begin{array}{ll}
	0       & if \ {n = 1}\\
	\frac{\alpha^{local}_{sync}}{f} + (\frac{\kappa^{local}_{sync}}{f} + \beta^{local}_{sync})*(n-2) +  \theta^{local}_{sync}   & if \ {n = r, n \geq 2}\\
	\frac{\alpha^{node}_{sync}}{f} + (\frac{\kappa^{node}_{sync}}{f} + \beta^{local}_{sync})*(n-2) +  \theta^{node}_{sync}      & otherwise.
\end{array} \right. 
\end{align}
$n=r$ means only one node is allocated for the job. $\alpha^{local}_{sync}$, $\beta^{local}_{sync}$
, $\theta^{local}_{sync}$, and $\kappa^{local}_{sync}$ are the parameters for the case when the job uses multiple GPUs of the same node.
$\alpha^{node}_{sync}$, $\beta^{node}_{sync}$
, $\theta^{node}_{sync}$, and $\kappa^{node}_{sync}$ are the parameters for the case when the job uses multiple nodes.

\parabf{Combining $T_{IO}$, $T_{grad}$, and $T_{sync}$.}
It is common for DL frameworks to overlap $T_{IO}$, $T_{grad}$, and $T_{sync}$ by pipelining storage IO, GPU computation, and network IO~\cite{JiangZLYCG20, MohanPKC22, PengZCBYLWG19, paszke2019pytorch}. We follow previous work~\cite{QiaoCSNH0GX21} to estimate $T_{iter}$ as somewhere between no overlap (i.e., $T_{IO} + T_{grad} +  T_{sync}$) and fully overlap (i.e., $max( T_{IO}, T_{grad},  T_{sync})$). Therefore, $T_{iter}$ is modeled as:
\begin{align} 
	T_{iter} = ((T_{IO}^{\gamma_1} + T_{grad}^{\gamma_1})^{\gamma_2/\gamma_1}+T_{sync}^{\gamma_2})^{\gamma_2}, \gamma_1 \geq 1, \gamma_2 \geq 1.
\end{align} 
$\gamma_1$ and $\gamma_2$ are fittable parameters. When $\gamma_1 = 1$ and $\gamma_2 = 1$, $T_{iter} = T_{IO} + T_{grad} +  T_{sync}$ and when $\gamma_1 \to \infty$ and $\gamma_2 \to \infty$, $T_{IO}$, $T_{grad}$, and $T_{sync}$ tend to be fully overlapped.

Figure~\ref{fig:step_time_model} shows how our throughput model is fitted to measured throughput results on different DNN models and different configurations. We measured the throughputs on real NVIDIA V100 GPUs. As is shown in the figures, The fitted models represent the real throughput performance well. We will evaluate the fitted models in detail in Section~\ref{sec:exp:model}.

\subsection{Modeling Energy Consumption}
\label{sec:model:e}

\begin{figure}
	\centering
	\subfloat[ResNet18.]{
		\begin{minipage}[b]{0.5\linewidth}
			\includegraphics[width=1\linewidth]{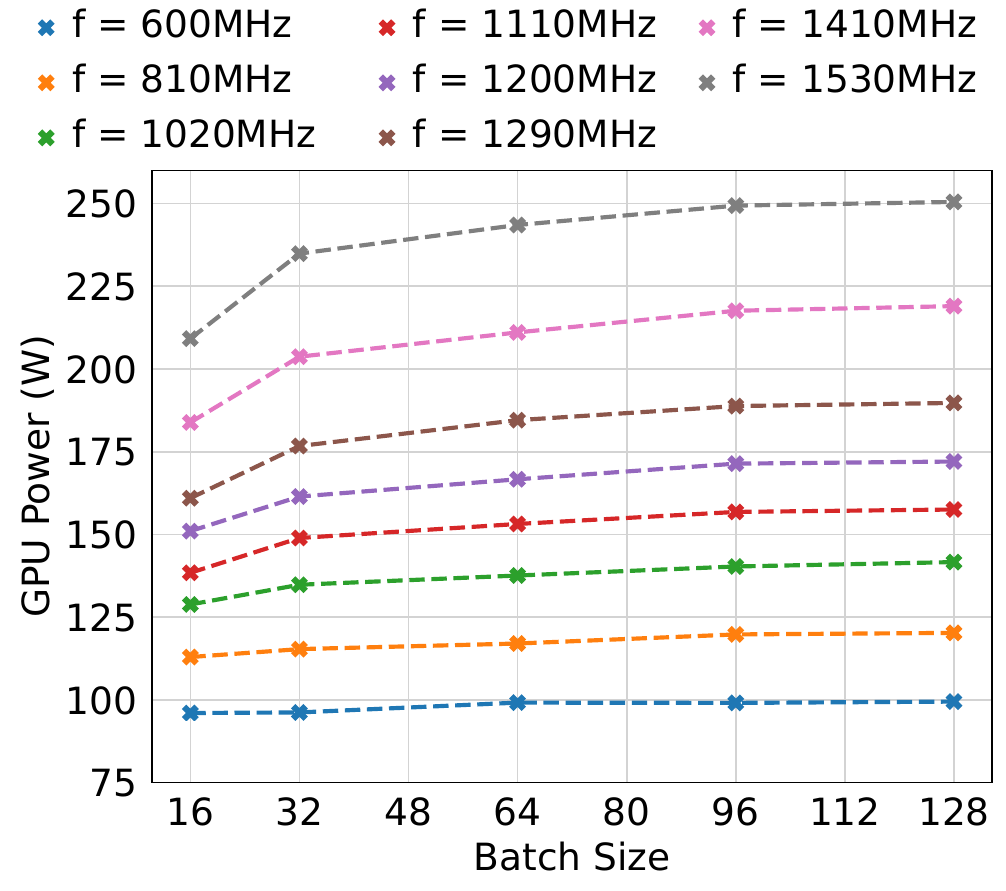}
		\end{minipage}
	}
	\subfloat[Deep Speech 2.]{
		\begin{minipage}[b]{0.5\linewidth}
			\includegraphics[width=1\linewidth]{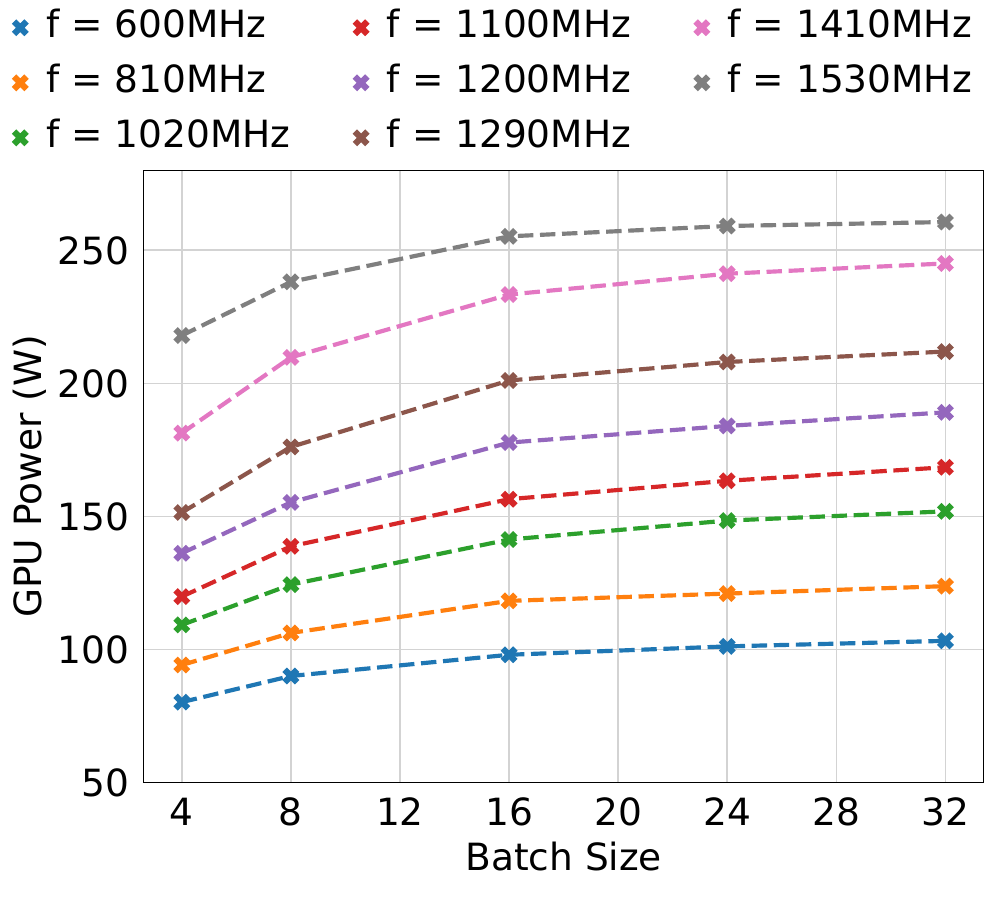}
		\end{minipage}
	}
	%\vspace{-0.1in}
	\caption{GPU power of training DNN models on one GPU.}
	%\vspace{-0.05in}
	\label{fig:p_bs}
\end{figure}

\begin{figure*}
	\centering
	\subfloat[ResNet18]{
		\begin{minipage}[b]{0.19\linewidth}
			\includegraphics[width=0.98\linewidth]{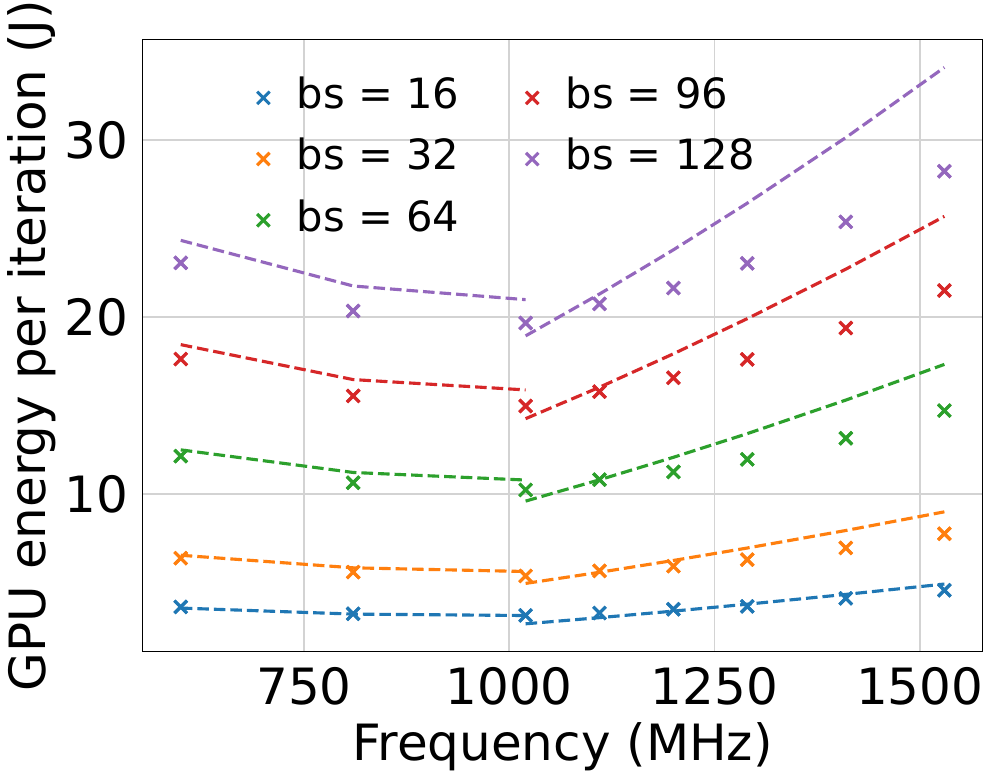} \\
			\includegraphics[width=0.98\linewidth]{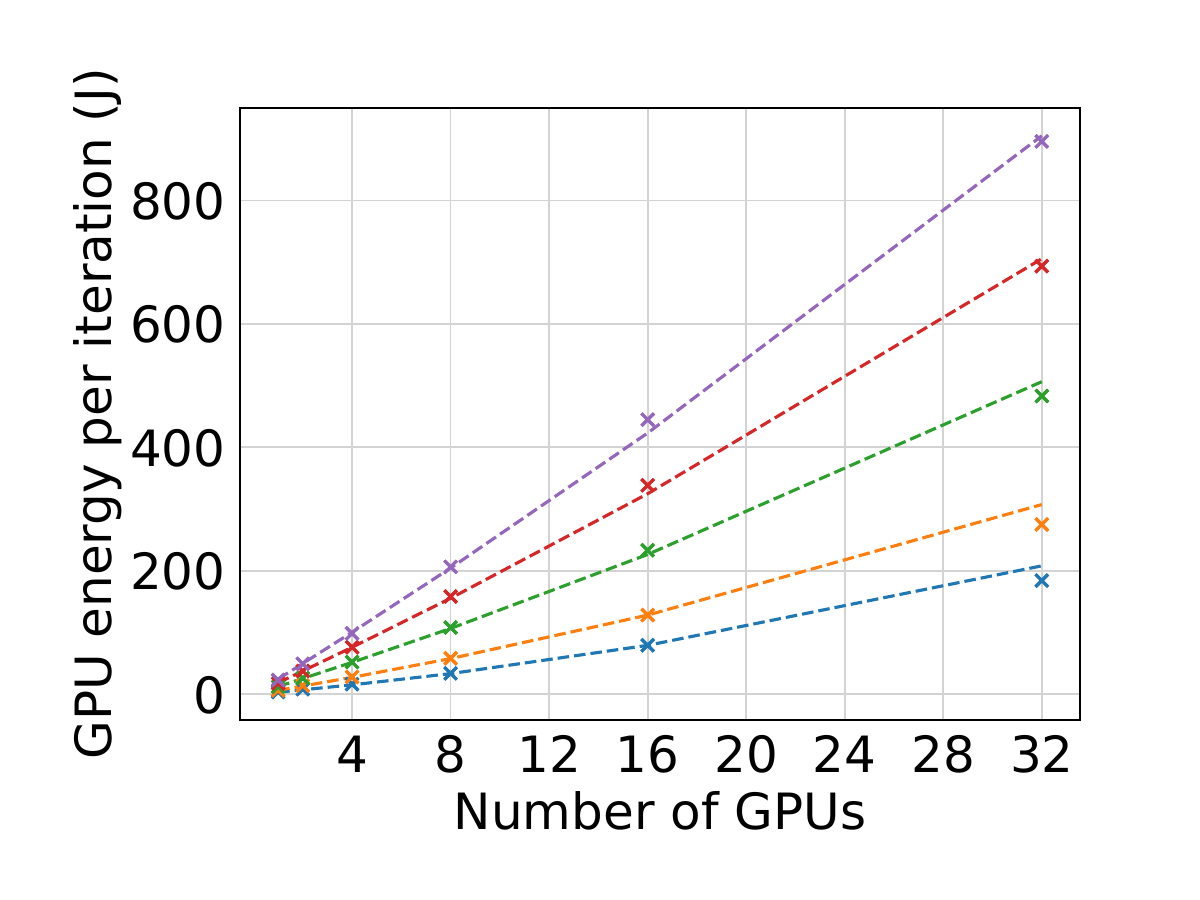} \\
			\includegraphics[width=0.98\linewidth]{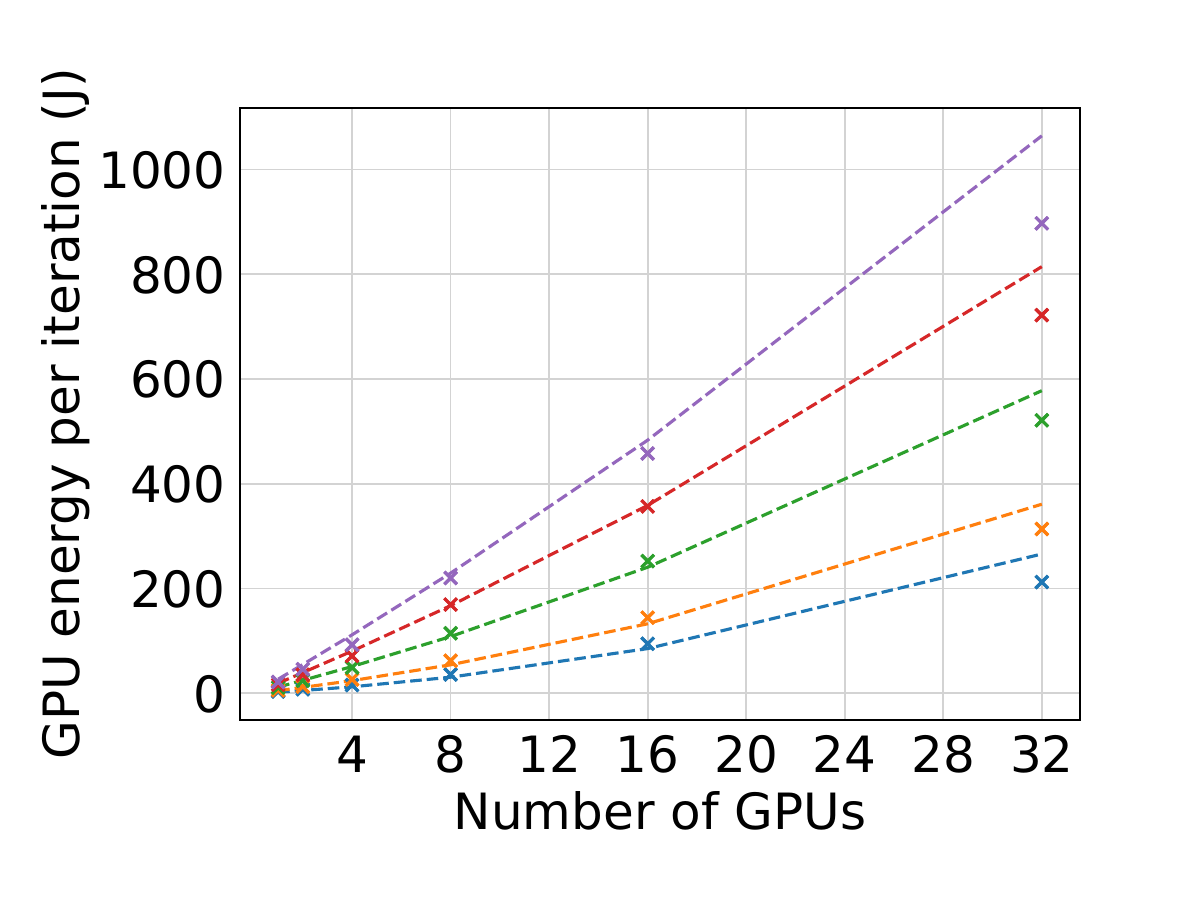}
		\end{minipage} 
	}
	\subfloat[VGG16]{
		\begin{minipage}[b]{0.19\linewidth}
			\includegraphics[width=1\linewidth]{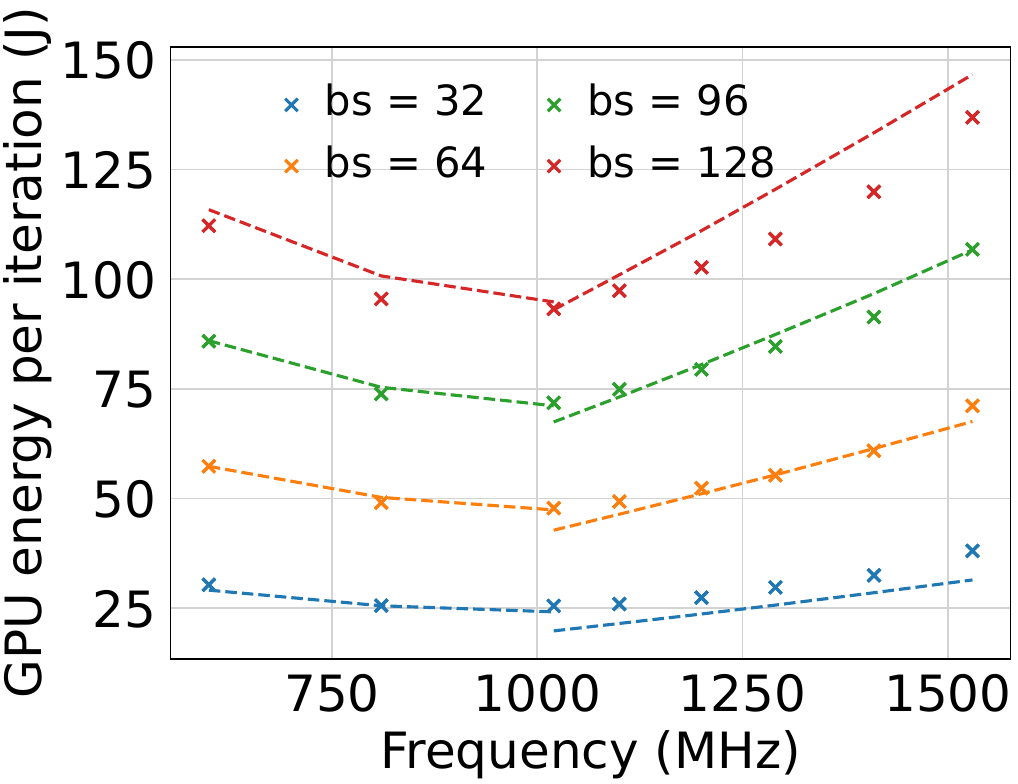} \\
			\includegraphics[width=1\linewidth]{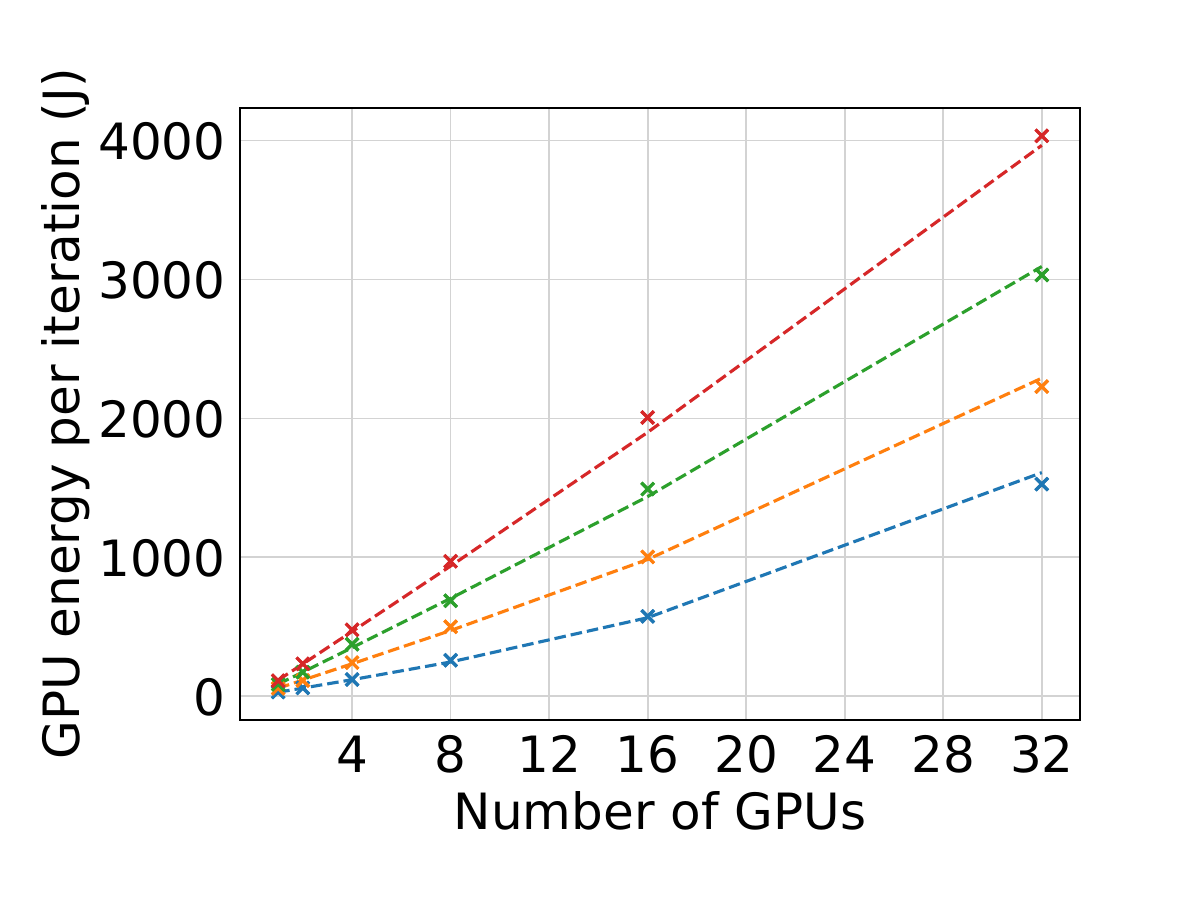} \\
			\includegraphics[width=1\linewidth]{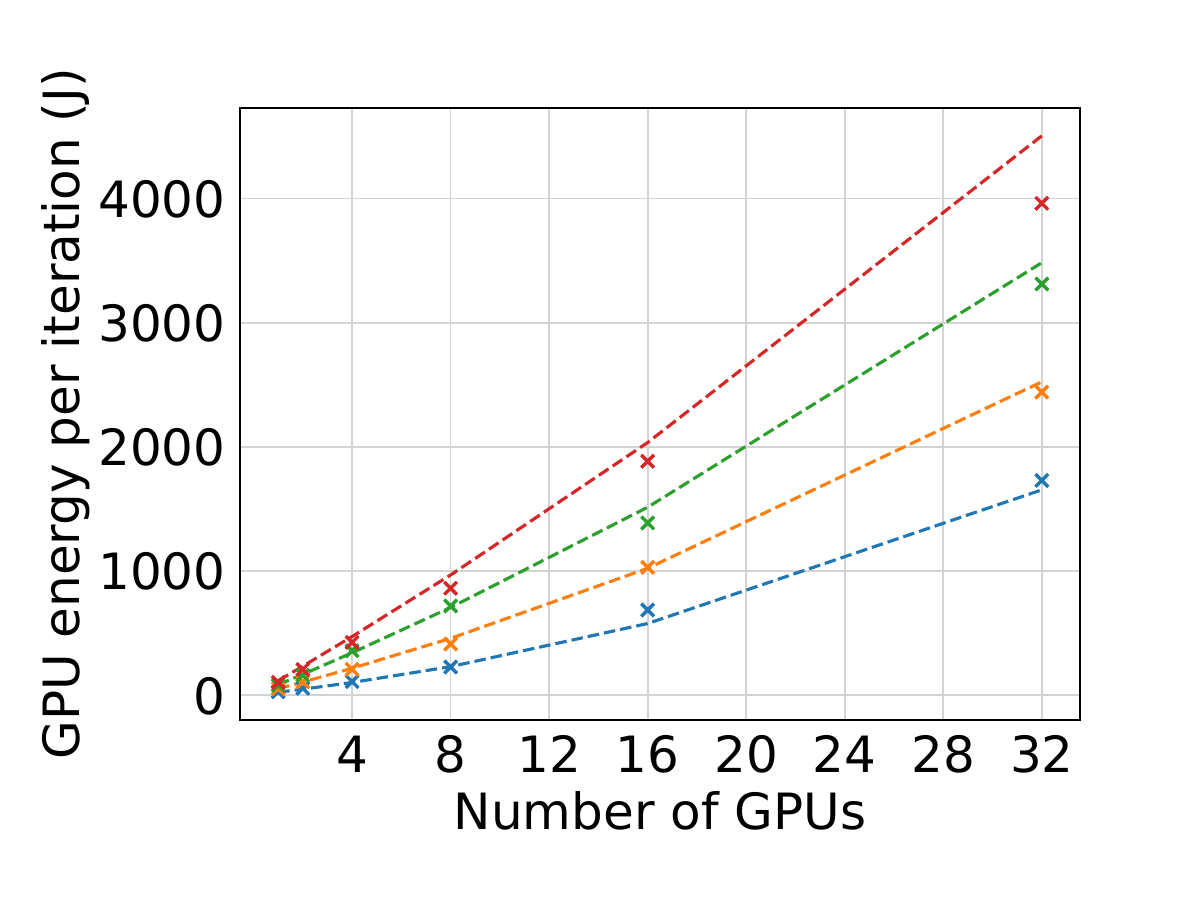}
		\end{minipage} 
	}
	\subfloat[Inception V3]{
		\begin{minipage}[b]{0.19\linewidth}
			\includegraphics[width=1\linewidth]{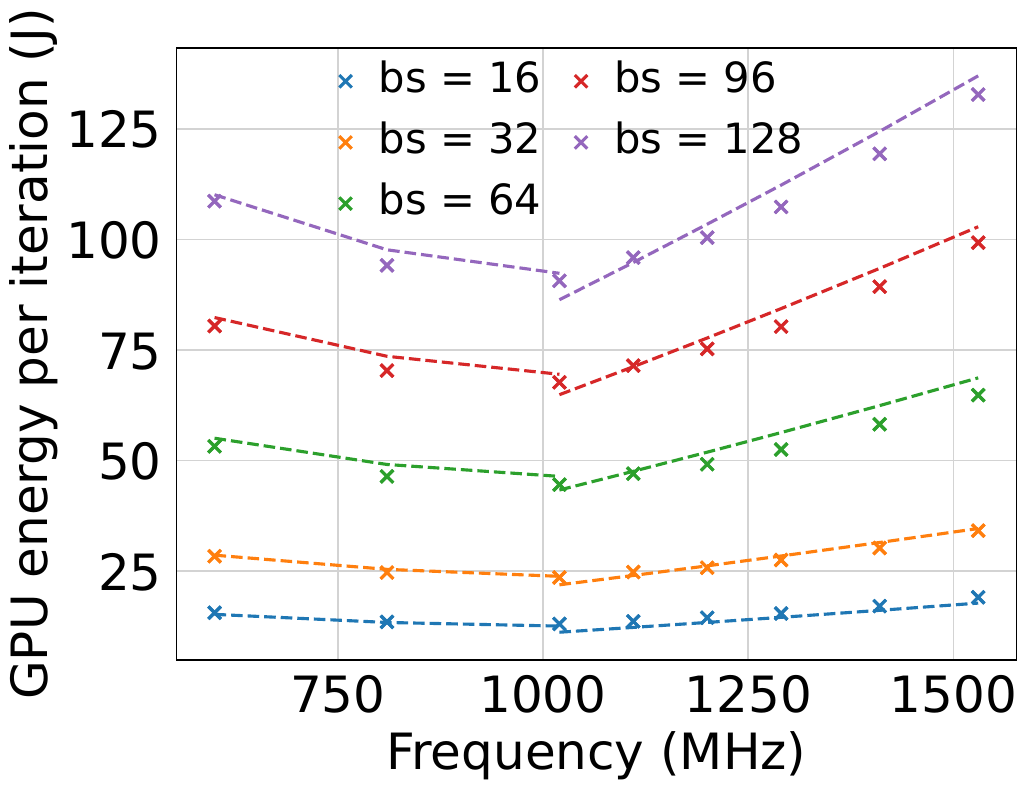} \\
			\includegraphics[width=1\linewidth]{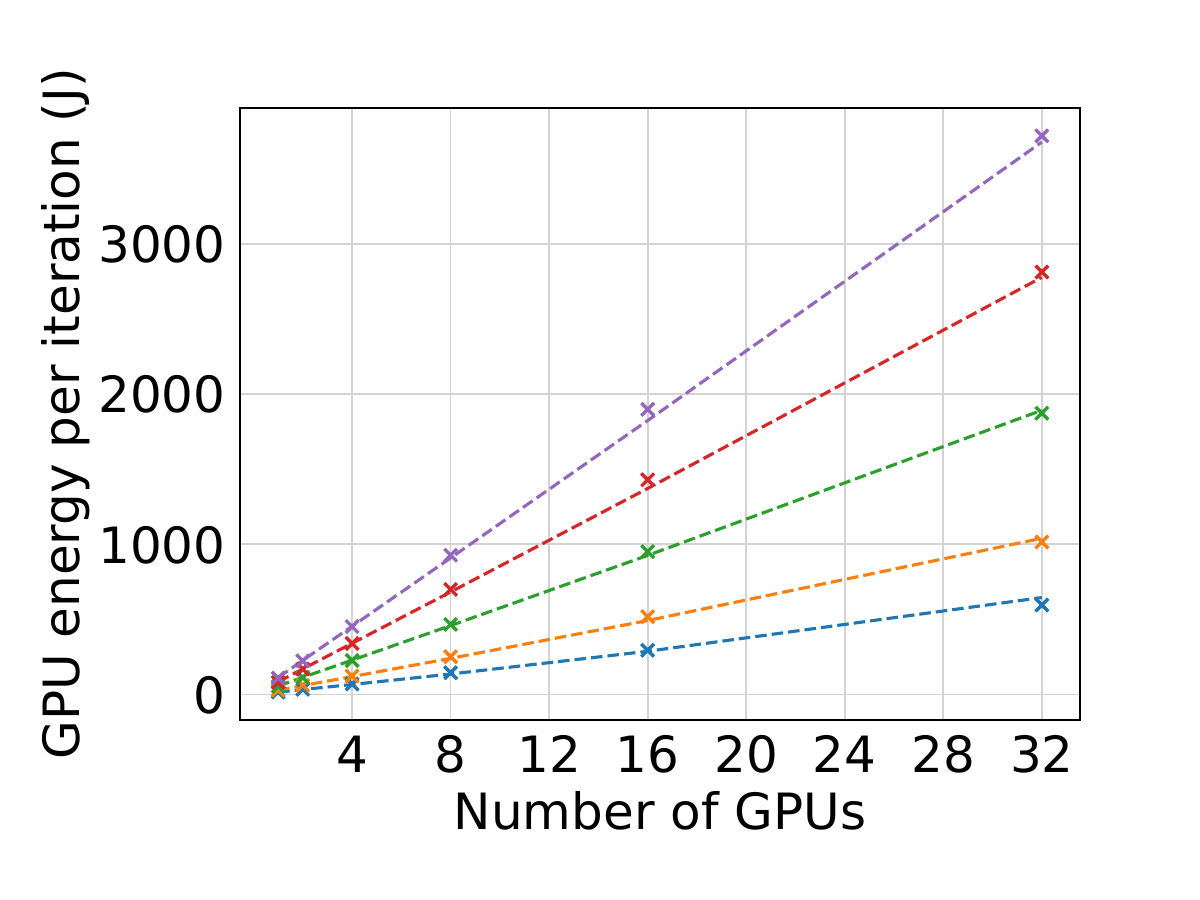} \\
			\includegraphics[width=1\linewidth]{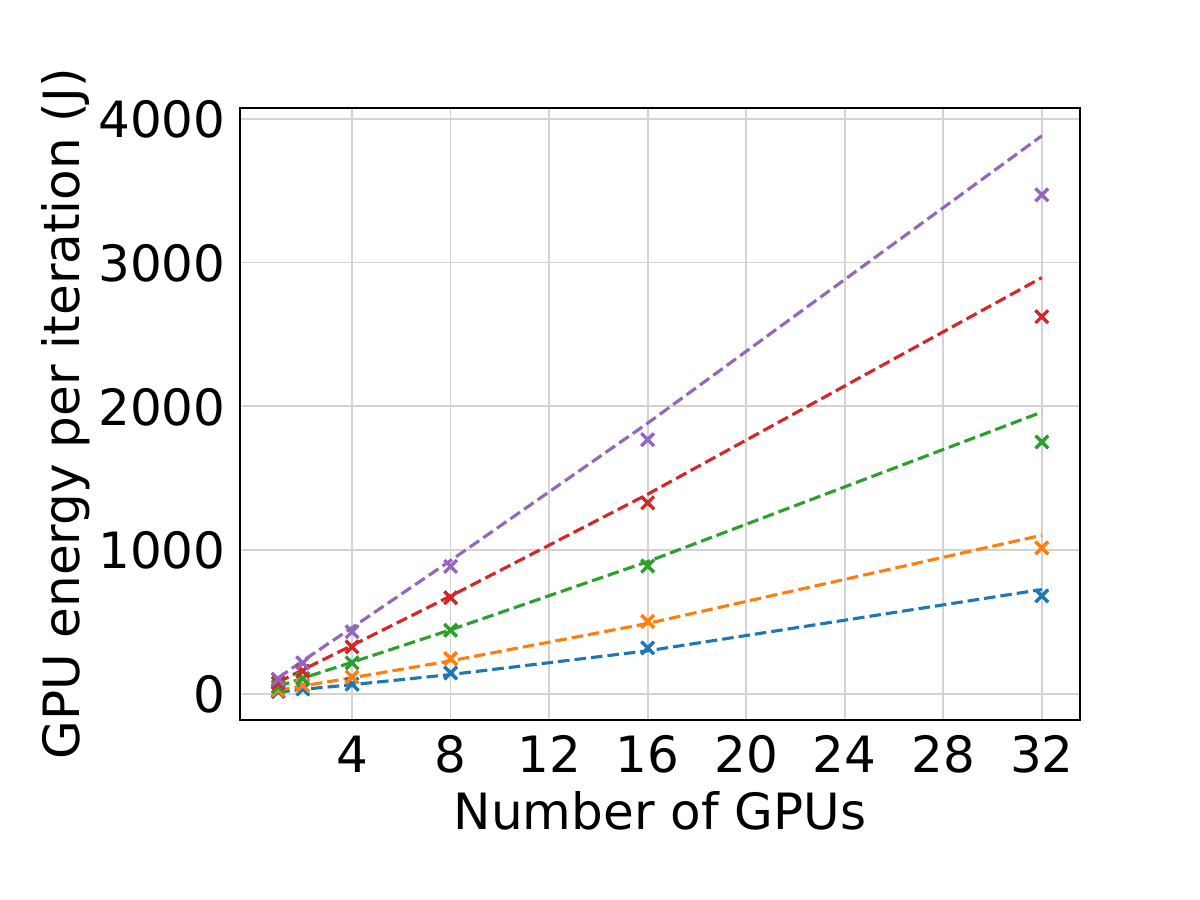}
		\end{minipage} 
	}
	\subfloat[GPT2]{
		\begin{minipage}[b]{0.19\linewidth}
			\includegraphics[width=1\linewidth]{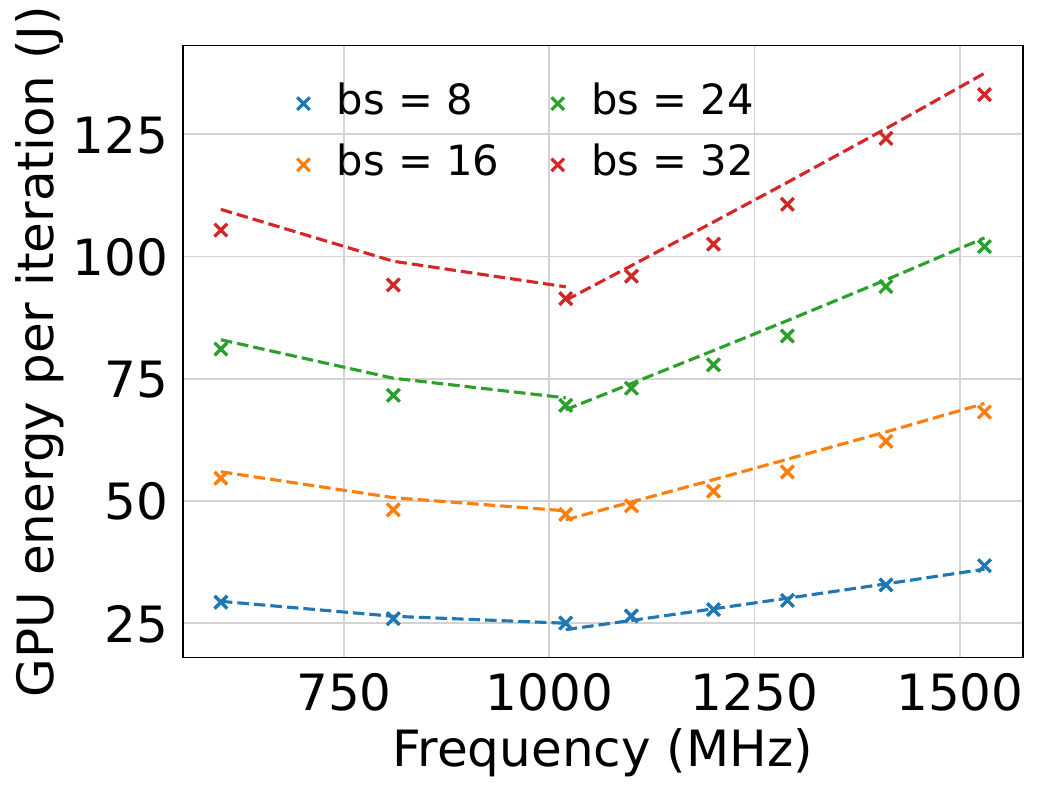} \\
			\includegraphics[width=1\linewidth]{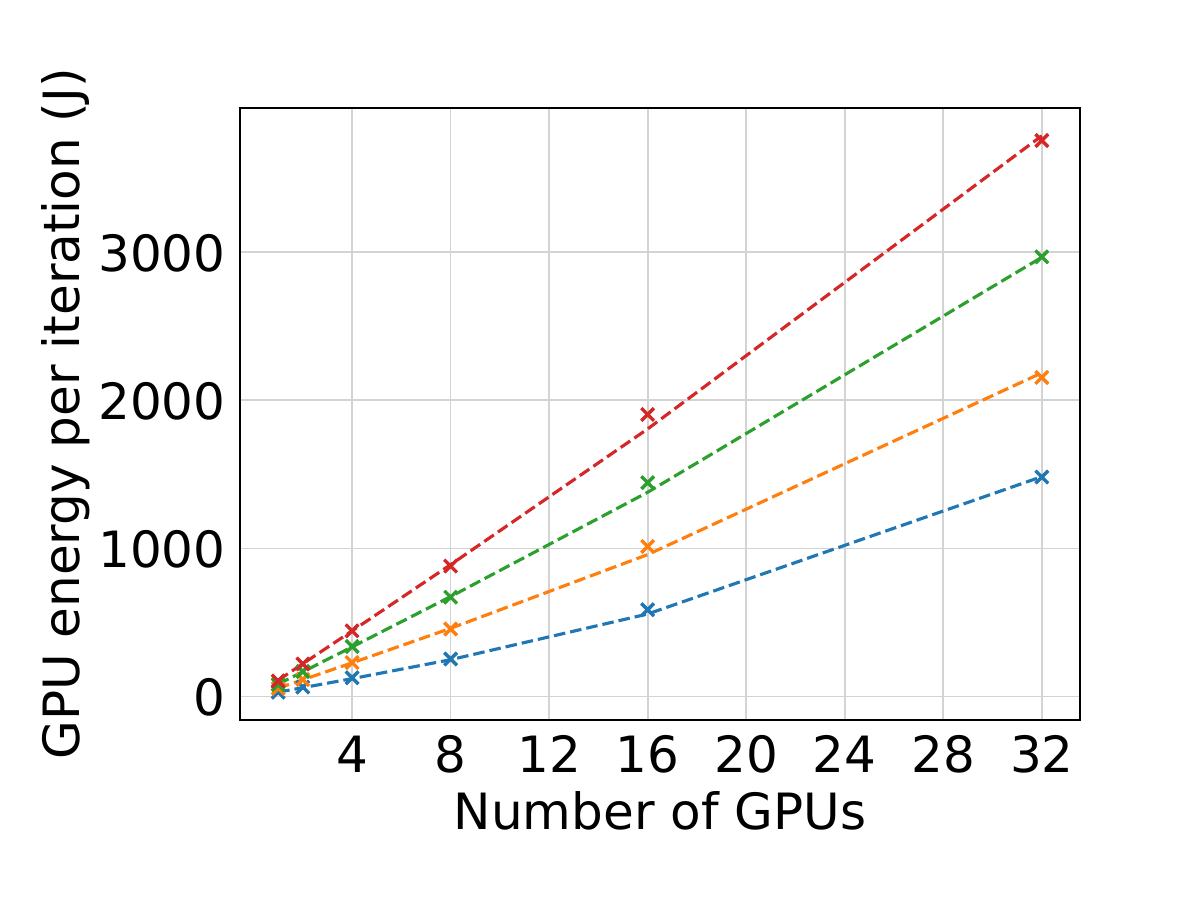} \\
			\includegraphics[width=1\linewidth]{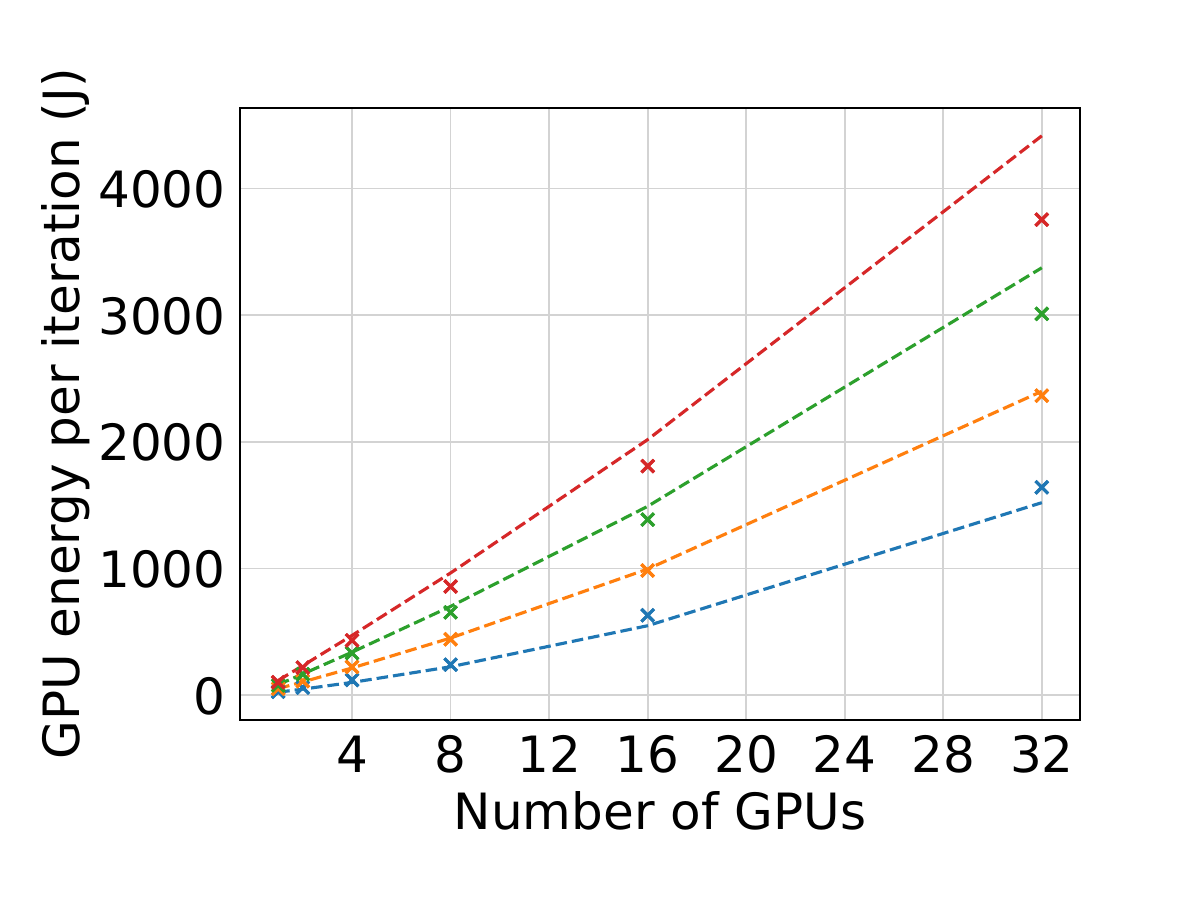}
		\end{minipage} 
	}
	\subfloat[Deep Speech 2]{
		\begin{minipage}[b]{0.19\linewidth}
			\includegraphics[width=1\linewidth]{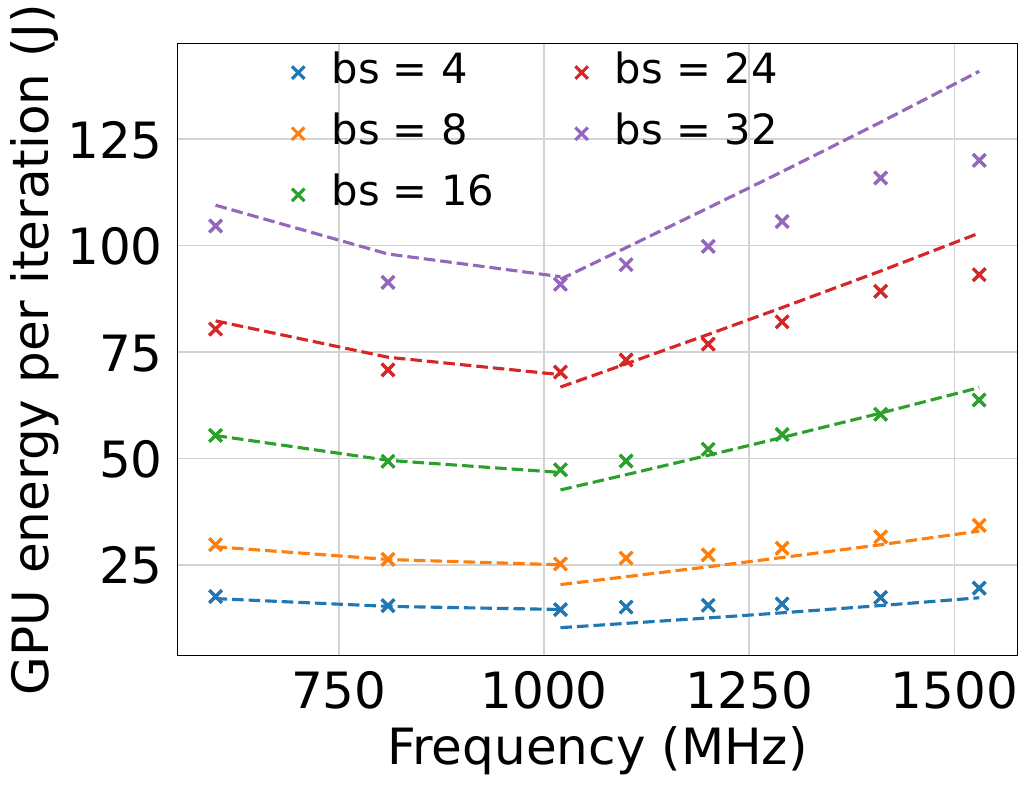} \\
			\includegraphics[width=1\linewidth]{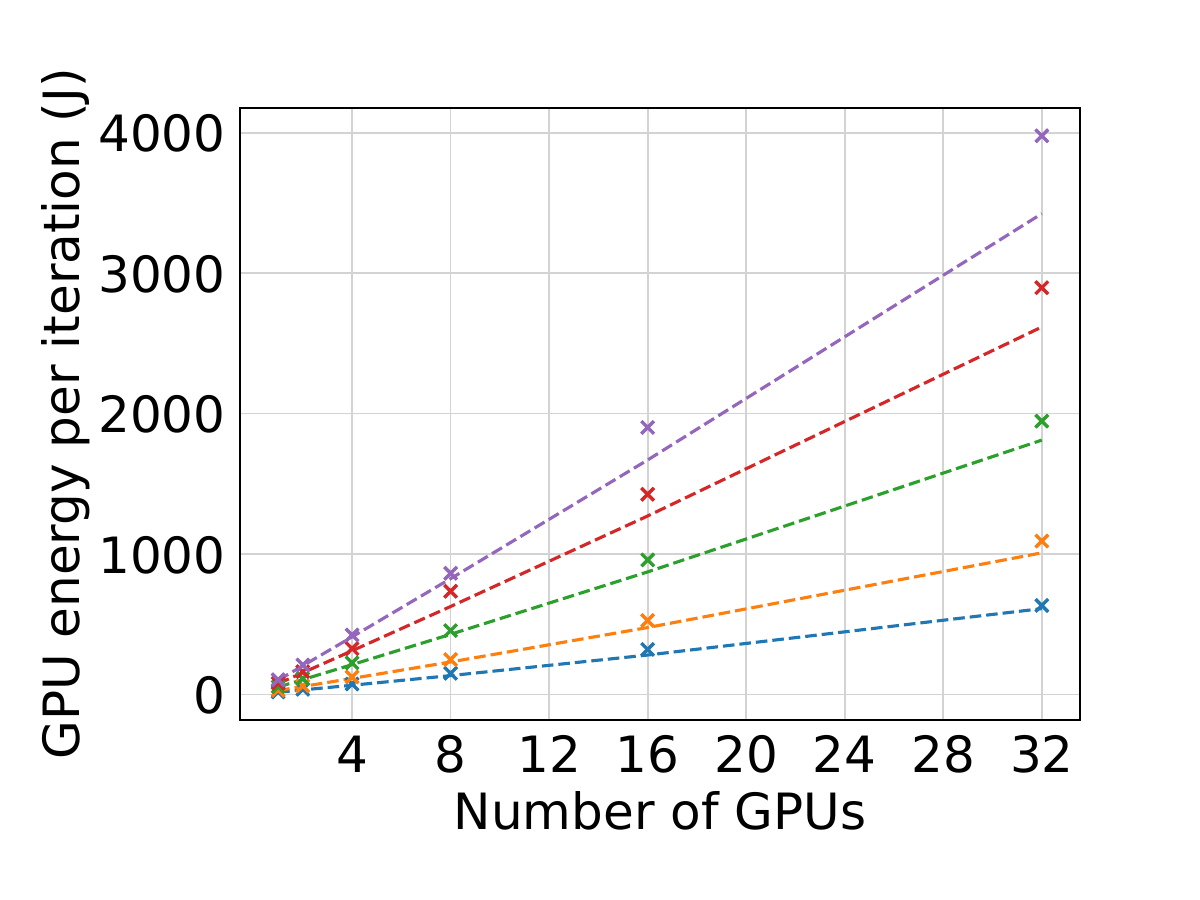} \\
			\includegraphics[width=1\linewidth]{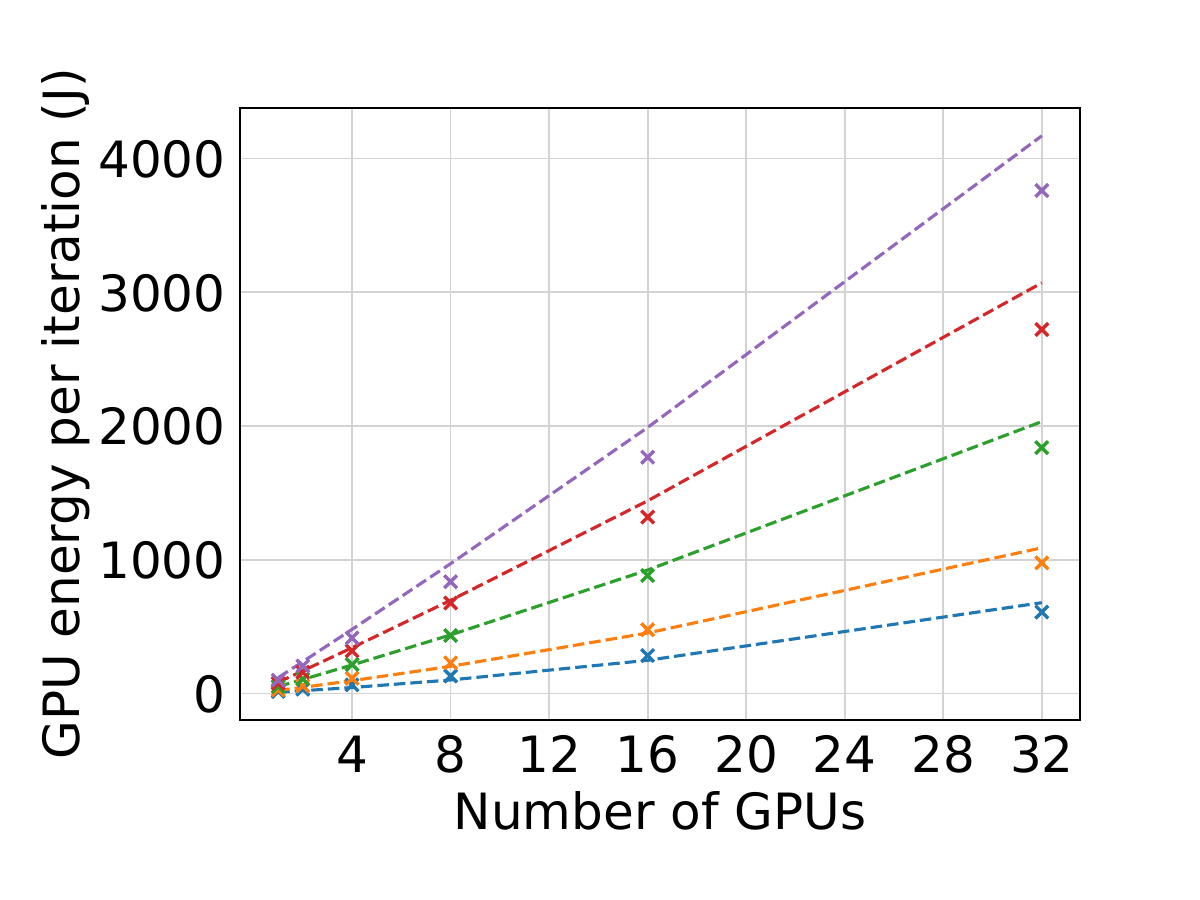}
		\end{minipage} 
	}
	%\vspace{-0.1in}
	\caption{GPU energy consumed per iteration by different DNN models with configurations. The scattered data points are the GPU energy values measured on real NVIDIA V100 GPUs and the lines show the predicted values of fitted models. The figures on the top show how GPU energy consumed per iteration changes with GPU frequency on one GPU. The figures in the middle show how GPU energy consumed per iteration changes with the number of GPUs with \SI{600}{MHz} frequency (low frequency). The figures at the bottom show how GPU energy consumed per iteration changes with the number of GPUs with \SI{1200}{MHz} frequency (high frequency). Note that the local batch sizes are kept the same and the global batch size scales linearly with the number of GPUs.}
	\vspace{-0.15in}
	\label{fig:energy_model}
\end{figure*}
The energy consumption of a GPU device can be decomposed into the sum of the dynamic and static components~\cite{GuerreiroIRT18}.
GPU devices take part in both the computation of gradients and the synchronization between GPUs, which are dynamic processes. Therefore, we model $E_{iter}$ as:
\begin{align} 
	E_{iter} = E_{grad} + E_{sync} + E_{static}.
\end{align} 

Then, we model the GPU energy consumption of different components in each iteration by modeling the average power of executing one iteration:
\begin{align} 
	E_{grad} = P_{grad} * T_{grad} * n; \\
	E_{sync} = P_{sync} * T_{sync} * n; \\
	E_{static} = P_{static} * T_{iter} * n,
\end{align} 
where $P_{grad}$, $P_{sync}$, and $P_{static}$ are the power (i.e., the consumed energy per unit of time) of different components on a single GPU. 
Previous studies~\cite{GonzalezGH97, butts2000static} proposed that the power of static and dynamic components can be modeled as:

\begin{align} 
	\label{eq:dynamic} P_{dynamic} = a * m * V^2 * f;\\
	\label{eq:static} P_{static} = V * N * K_{design} * \tilde{I}_{leak},
\end{align} 
where $a$ denotes the average utilization ratio, $C$ is the total capacitance, $V$ is the supply voltage. $N$, $K_{design}$, and $\tilde{I}_{leak}$ are constant parameters determined by the hardware.  Therefore, in these two models, the power of dynamic and static components scales with the frequency $f$ and the voltage $V$. 

Although it is common for GPU manufacturers to provide tools to get the frequencies of GPU devices, there is no easy way to know the voltage directly, nor how voltage scales with frequency~\cite{GuerreiroIRT18}. Therefore, to model the energy consumption of DL training jobs, we first model how the voltage of a GPU scales with frequency.

\parabf{GPU Voltage and frequency.} 
The voltage of a GPU scales with GPU frequency in a piecewise function: the voltage is a constant value when the frequency is low; after a specific frequency, the voltage increases linearly with the frequency~\cite{GuerreiroIRT18}. This specific frequency depends on the hardware device. Therefore, according to Equation~\ref{eq:dynamic}, $P_{grad}$ and $P_{sync}$ scale linearly with $f$ at lower frequencies and is a cubic complements of $f$ at higher frequencies; according to Equation~\ref{eq:static}, $P_{static}$ is constant at lower frequencies and is a linear function of $f$ at higher frequencies.
%This is in accordance to the measured GPU power that we measured on real V100 GPUs: as is shown in Figure~\ref{x}, for different batch sizes, the GPU power is a linear function to GPU frequency when $f < 1000$ and is a cubic function to GPU frequency when $f >= 1000$. 
Next, we separately model $P_{grad}$ , $P_{sync}$, and $P_{static}$ for low frequencies and high frequencies.

\parabf{Modeling $P_{grad}$.} 
Except for GPU frequency, $P_{grad}$ is also related to the local batch size on each GPU. The larger the local batch size, the larger the power of a GPU. However, the power of a GPU does not scale linearly with local batch size. As is shown in Figure~\ref{fig:p_bs}, we observe that the power of the GPU scales sublinearly with local batch size. Therefore, we model $P_{grad}$ as a logarithmic function of $bs$:

\begin{align} 
	\small
	P_{grad}=\left\{
	\begin{array}{ll}
		\kappa_{grad} * (\alpha_{l_{g}} * log(bs + \theta_{l_{g}} + \beta_{l_{g}}))  & if \ {f < f_0}\\
		\kappa_{grad} * (\alpha_{h_{g}} * log(bs + \theta_{h_{g}} + \beta_{h_{g}}))    & otherwise;
	\end{array} \right. 
\end{align}
\begin{align} 
	\small
	\kappa_{grad}=\left\{
	\begin{array}{ll}
	a_{l_{g}} * f + b_{l_{g}} & if \ {f < f_0}\\
	a_{h_{g}} * f^3 +   b_{h_{g}} * f^2 + c_{h_{g}} * f + d_{h_{g}}   & otherwise,
	\end{array} \right. 
\end{align}
where $\alpha$, $\beta$, $\theta$, $a$, $b$, $c$, $d$ are learnable parameters, $f_0$ is the hardware-dependent breaking point between low frequencies and high frequencies.

\parabf{Modeling $P_{sync}$.} 
Similar to $P_{grad}$, we model $P_{sync}$ at both low and high frequencies. The time spent on model parameters synchronization is only relevant to the DNN model and the bandwidth between GPU devices. So $P_{sync}$  is modeled as:

\begin{align} 
	\small
	P_{sync}=\left\{
	\begin{array}{ll}
		a_{l_{s}} * f + b_{l_{s}} & if \ {f < f_0}\\
		a_{h_{s}} * f^3 +   b_{h_{s}} * f^2 + c_{h_{s}} * f + d_{h_{s}}   & otherwise,
	\end{array} \right. 
\end{align}
where $a$, $b$, $c$, $d$ are learnable parameters, $f_0$ is the hardware-dependent breaking point between low frequencies and high frequencies.

\parabf{Modeling $P_{static}$.} 
According to Equation~\ref{eq:static}, $P_{static}$ is proportional to $V$:
\begin{align} 
	P_{static}=\left\{
	\begin{array}{ll}
		P_{static_{l}} & if \ {f < f_0}\\
		c_h * f   & otherwise,
	\end{array} \right. 
\end{align}
where $P_{static_{l}}$ and $ c_h$ are parameters that can be fitted.

Figure~\ref{fig:energy_model} shows how our energy model is fitted to measured energy consumption results on different DNN models and different configurations. We measured the throughputs on real NVIDIA V100 GPUs. As is shown in the figures, The fitted models represent the real energy consumption performance well. We will evaluate the fitted models in detail in Section~\ref{sec:exp:model}.

\subsection{Energy-throughput Tradeoffs}
\begin{figure}[t]
	\centering
	\includegraphics[width=0.85\linewidth]{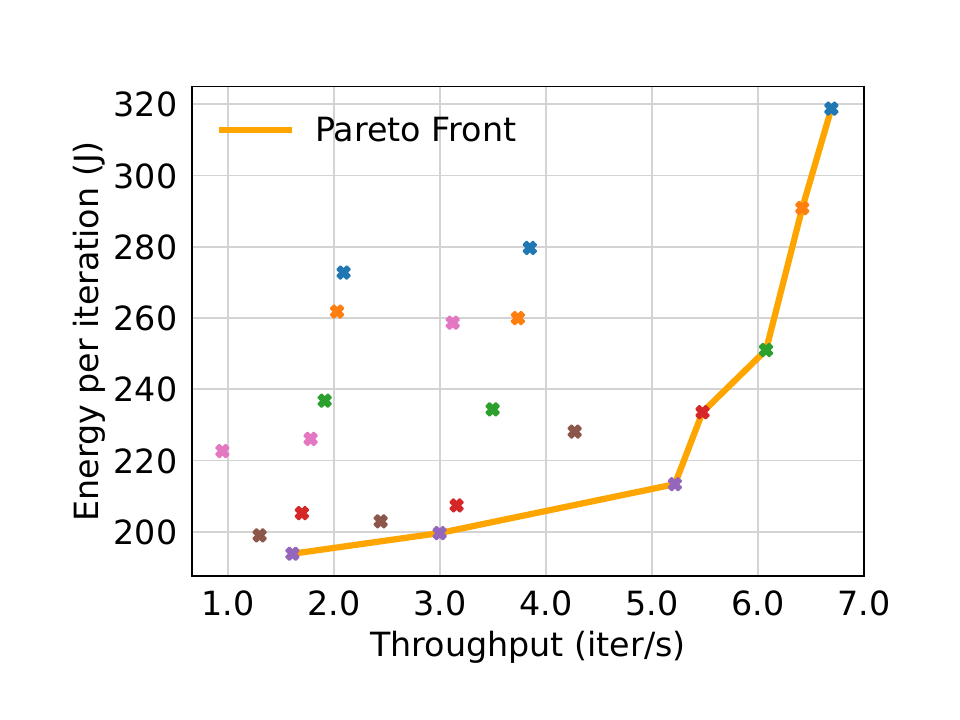}
	\caption{The energy-throughput tradeoff. The data points in the same color represents the values measured under the same GPU frequencies.}
	\label{fig:e_jct}
	%\vspace{-0.1in}
\end{figure}
It would be ideal to achieve a small JCT with little energy consumption. 
However, from our throughput model (Section~\ref{sec:model:tpt}), we can see that if we want to improve the throughput of a DL training job while keeping the global batch size unchanged, we need to use more GPU devices and set the GPUs to the highest supported frequency. However, according to our energy-consumption model (Section~\ref{sec:model:e}), training a DL job with a large number of GPUs and the highest GPU frequency consumes more energy than many other settings. There is a tradeoff between the JCT and the energy consumption of a DL training job.

Figure~\ref{fig:e_jct} is an example to show this tradeoff. The values were measured by training GPT2 with a global batch size of 64 on real NVIDIA V100 GPUs. Each data point represents the throughput and the energy consumed per iteration of one feasible configuration. The data points on the orange curve achieve  Pareto optimality~\cite{censor1977pareto}, for which we cannot reduce energy consumed per iteration without sacrificing throughput, and vice versa. Training with the configurations that are not on this curve would not be \textit{energy-efficient}, as there are configurations that can achieve higher throughputs with similar energy consumption, or consume less energy with a similar throughput.

For a single DL training job, we can find a configuration that achieves Pareto equilibrium by constantly trying to use a different number of GPUs and GPU frequency.
However, it is more complicated to achieve the tradeoff in a GPU cluster where the total number of GPUs is limited. If one training job runs under a configuration that achieves Pareto equilibrium, there might not be enough GPUs left for other jobs to achieve  Pareto equilibrium. Also, with an energy budget, there is no straightforward solution to decide which job should use more GPUs or which GPU's frequency should be higher.

	\section{System Design}
\subsection{System Overview}
\begin{figure}[t]
	\centering
	\includegraphics[width=1\linewidth]{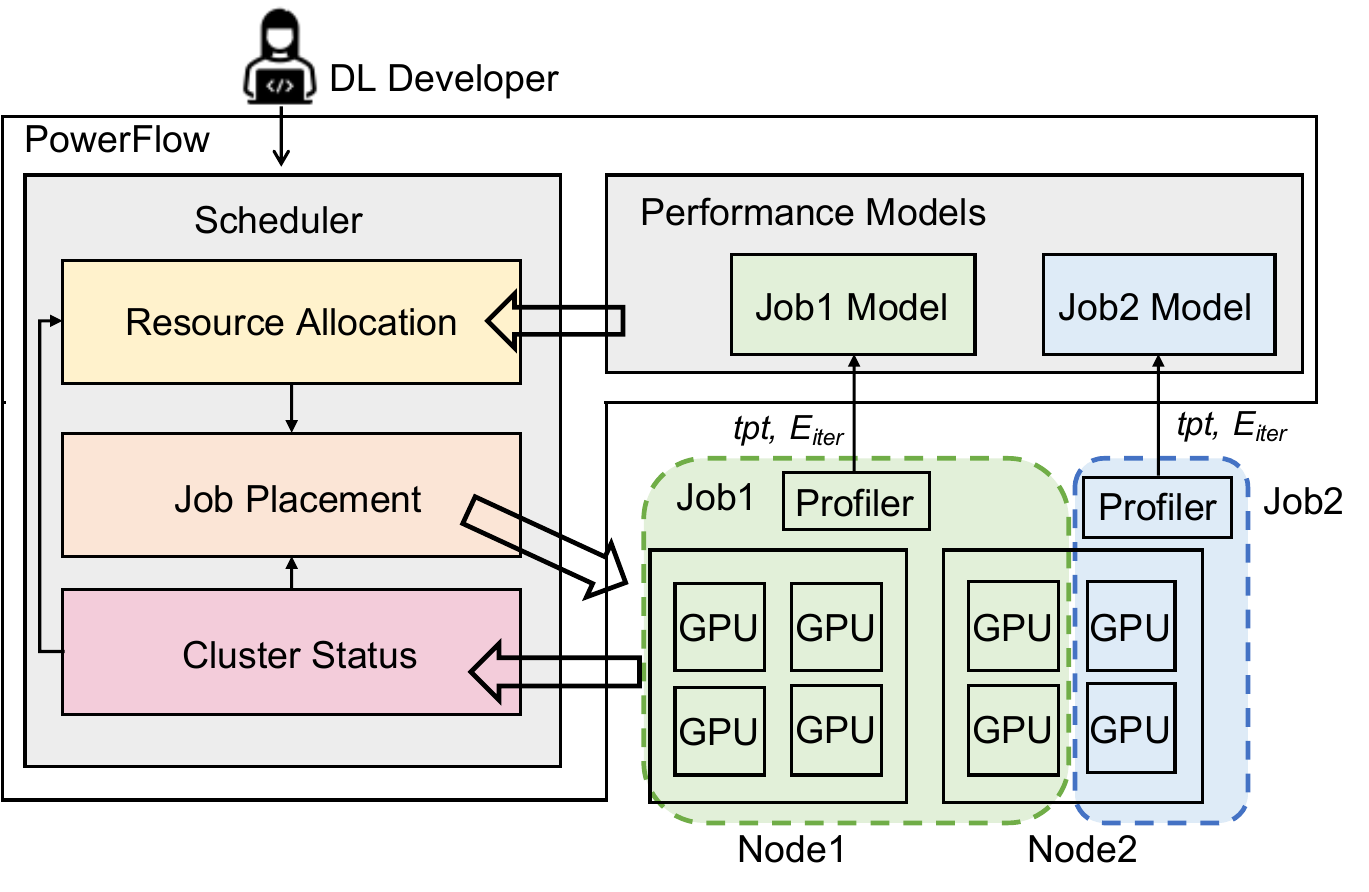}
	\caption{\sysname{} architecture.}
	\label{fig:overview}
	%\vspace{-0.1in}
\end{figure}

 Figure~\ref{fig:overview} shows the architecture of \sysname{}. DL developers submit DL training jobs to  \sysname{}. For each job, a throughput model and an energy consumption model are built to predict the DL training job's performance under different configurations (i.e., the number of GPUs used and the GPU frequency). To make more accurate performance predictions, \sysname{} first pre-runs the submitted jobs offline on one GPU with different frequencies and profiles the performance to fit the performance models. 
 This takes about four minutes. As DL training jobs usually run for hours or even days, this pre-running overhead is rather small and has little influence on the final average JCT.
 Then, \sysname{} schedules the jobs with a resource allocation module and a job placement module.
 
The resource allocation module decides the number of GPUs allocated to each job based on its performance models.
Upon each scheduling event such as job arrival or completion, the resource allocation module may update the resource allocation of some jobs through elastic scaling, i.e., adjusting the number of GPUs allocated to a job based on the job's real performance and the number of available GPUs. The module also computes the local batch size (i.e., dividing global batch size by the number of GPUs) for a given job. The job placement module selects GPUs from the cluster for each job based on the topology. While each job is running, \sysname{} profiles the real throughput and energy consumption of each iteration. Based on the online and offline profiled performance, \sysname{} continuously fits the performance models of each job to make more accurate performance predictions and better scheduling decisions.

\subsection{Resource Allocation}
\label{sec:design:algo}
\begin{algorithm}[t]
	\caption{\sysname{} Resource Allocation}
	\begin{algorithmic}[1]
		%\Function{GetPriorityG}{job}
		%\State $\Delta_{jct} \gets job.jct(job.n, job.f) - job.jct(job.n + 1, job.f)$
		%\State $\Delta_E \gets job.E(job.n, job.f) - job.E(job.n + 1, job.f)$
		%\State $relative\_jct = \Delta_{jct} / JCT$
		%\State $relative\_e = \Delta_{e} / E$
		%\State $job.priority_g = relative\_jct / relative\_e$
		%\EndFunction
		%\Function{GetPriorityF}{job}
		%\State $\Delta_{jct} \gets job.jct(job.n, job.f) - job.jct(job.n, job.f + \Delta_f)$
		%\State $\Delta_E \gets job.E(job.n, job.f) - job.E(job.n, job.f + \Delta_f)$
		%\State $relative\_jct = \Delta_{jct} / JCT$
		%\State $relative\_e = \Delta_{e} / E$
		%\State $job.priority_f = relative\_jct / relative\_e$
		%\EndFunction
		\Function{\sysname}{jobs}
		\State \textbf{ // GPU allocation}
		\For{$i$ from $0$ to $J$}
		\State $i.n \gets 0$
		\State \textbf{ // The most energy efficient GPU frequency}
		\State $i.f \gets f_0$
		\State $i.priority_g \gets$ \Call{GetPriorityG}{$i$}
		\State $QueueG.insert(i)$
		\EndFor
		%\State $E^\ast \gets E$
		\State $g \gets G$
		\While{$g > 0$}
		\State \textbf{ // Pick the job with largest marginal return}
		\State $i \gets QueueG.dequeue()$
		%\State $E^\ast \gets E^\ast - i.E(job.n) + i.E(i.n + 1)$
		\State $i.n \gets i.n + 1$
		\State $g \gets g - 1$
		\State \textbf{// Update the marginal return of job $i$}
		\State $i.priority_g \gets$ \Call{GetPriorityG}{$i$}
		\State $QueueG.insert(i)$
		\If{$g \parallel gpu\_per\_node = 0$}
		    \If{$\sum{j=0}^{J}i.P \ge \eta * G * P_{max}$}
		        \State \textbf{Break}
		    \EndIf
		\EndIf
		\EndWhile
			\State \textbf{ // GPU frequency configuration}
		\For{$i$ from $0$ to $J$}
		\State $i.priority_f \gets$ \Call{GetPriorityF}{$i$}
		\State $QueueF.insert(i)$
		\EndFor
		\While{$\sum{j=0}^{J}i.P \le \eta * G * P_{max}$}
			\State \textbf{ // Pick the job with largest marginal return}
		\State $i \gets QueueF.dequeue()$
		\State $i.f \gets i.f + \Delta_f$
		\State $i.priority_g \gets$ \Call{GetPriorityF}{$i$}
		\State $QueueD.insert(i)$
			\EndWhile
		\EndFunction
	\end{algorithmic}
	\label{algo}
\end{algorithm}

In this section, we describe the detailed design of the resource allocation module which finds the tradeoff between JCT and energy consumption.
We propose \textit{energy efficiency} to capture whether the consumed energy brings enough throughput improvement, based on which \sysname{} dynamically allocates
resources to submitted jobs. \sysname{} applies network packing and
buddy allocation to avoid resource
fragmentation, thus reducing energy consumption on idle GPUs.

\parabf{Energy efficiency.}
The key to achieving a short JCT with an energy budget is to be more \textit{energy-efficient}. 
For a DL training job with a fixed global batch size, the larger the throughput, the smaller the JCT.
Therefore, we define \textit{energy efficiency} as the throughput brought by each unit of energy consumption.For job $j$, we have:
\begin{align} 
	\label{eq:ee_tpt}ee_j=\frac{tpt_j}{E_j},
\end{align}
where $ee$ represents the energy efficiency of a DL training job, $E$ is the total energy consumed to finish the job, and $tpt$ is the average throughput of the job. DL developers usually specify
a maximum number of iterations to execute as a termination condition of a DL training job. Let us denote this maximum iteration number of job $j$ as $iter_j$. Combining Equation~\ref{eq:tpt} and~\ref{eq:ee_tpt}, we have:
\begin{align} 
	\label{eq:ee}ee_j=\frac{iter_j}{JCT_j * E_j}.
\end{align}

According to our performance models in Section~\ref{sec:model:tpt} and Section~\ref{sec:model:e}, if the GPU frequency is too large, the total energy consumption would be large, thus is energy-inefficient; if the GPU frequency is too small, the JCT would be very long, leading to poor energy efficiency as well. 

\parabf{The energy-aware scheduling problem.} 
%Take the training job in Figure~\ref{fig:e_jct} as an example:
%the configurations correspond to the data points on the Pareto Front are more energy-efficient than other configurations, as these configurations achieves comparable throughputs with less energy consumed. 
%from figure, the data point on the pareto front; min GPU num is 
To tradeoff between average JCT and total energy consumption with an energy budget,
we propose a simple parameter $\eta  \in$ [0, 1): $\eta$ represents how much \textit{power} the service provider wishes to consume compared to the case where the cluster is fully allocated with the default frequency.
$\eta$ can converse to an ``energy budget'' by $budget = \eta * G * P_{max} * t$, where $G$ is the total number of GPUs in the cluster, $P_{max}$ is the average GPU power when executing a DL training job with the highest supported GPU frequency, and $t$ is the period of time during which the energy consumption in the cluster does not exceed $budget$.
\sysname{} allocates as many GPUs as possible to the jobs and uses the highest GPU frequencies when $\eta=1$ and allocates as few GPUs as possible with the most energy-efficient frequencies when $\eta$ gets closer to $0$. The smaller the $\eta$, the smaller the power of the cluster. The whole cluster may consume even more energy with a small $\eta$: although the energy consumed by the cluster in every $t$ does not exceed $budget$, the time needed to finish all submitted jobs might be multiplied several times, which is not energy-efficient.
Therefore, the cloud provider should choose a proper $\eta$.

%a naive way is to decrease frequency and use the least number of GPU. But this is not enough. if all use the configuraition with least e consumption, the jct is too large.
With $\eta$, \sysname{} allocates GPUs to submitted jobs while making sure that the total power of the cluster does not exceed $power\_limit = \eta * G * P_{max}$. The power of each job can be estimated by our performance model as $P = E_{iter} / T_{iter}$. Suppose that there are $J$ jobs submitted to the cluster, we try to solve the following optimization problem:
\begin{align} 
	\min~&~\frac{1}{n}~\sum_{j=0}^{J-1} JCT_j \\
	s.t.~&~\sum_{k=0}^{G-1} P_i  \leq \eta * G * P_{max}.
\end{align} 

A straw-man solution is to search for the most energy-efficient configuration of each job and execute the jobs with these energy-efficient configurations.
This makes the jobs energy-efficient if the cluster has enough GPUs to run all of the submitted jobs.
However, this solution does not consider the limited resources in a cluster: if a job uses more GPUs to achieve shorter JCT or uses low GPU frequency to reduce energy consumption, the other jobs in the cluster might have to wait for the GPU resources for a longer time. This might lead to an even longer average JCT compared to the schedulers that are not energy-aware.

From the straw-man solution, we can see that the key to achieving our goal is to consider energy efficiency from the perspective of the GPU cluster. 
We should not only consider how the performance would change if we adjust the configuration of a DL job, but also the influence on the whole cluster. 

Guided by this insight, \sysname{} starts by executing all of the jobs with the most energy-efficient GPU frequency. Then, \sysname's resource allocation module schedules the submitted jobs in two steps: (1) allocate the GPUs in the clusters one by one under the most energy-efficient frequencies;
(2) if the total power of the cluster does not exceed $power\_limit$, increase the GPU frequencies to further accelerate the jobs. Algorithm~\ref{algo} shows the pseudocode of these two steps, which we will describe separately in detail. 

\parabf{GPU allocation.} 
We develop a greedy algorithm to solve the GPU allocation problem (lines 10-20). The intuition is to allocate the GPUs to the job with the highest \textit{marginal return}. \sysname{} takes this marginal return as the \textit{priority} of each submitted job. 
The \textit{marginal return} of job $j$ for GPU allocation is defined as:
\begin{align} 
	\footnotesize
	priority_G = \frac{(JCT_{j,n,f} - JCT_{j,n+1,f})/JCT}{(E_{j,n+1,f} - E_{j,n,f})/E},
\end{align} 
where the numerator is the relative JCT reduction of allocating one more GPU to $j$ compared to the total JCT of the whole cluster, and the denominator is the relative energy increase compared to the total energy consumption of the cluster.

\sysname{} maintains a priority
queue to order the jobs in $priority_G$ (lines 12-17). 
For each iteration, the algorithm dequeues the head from the
queue, which is the job with the largest marginal return (lines 11-12). The
algorithm allocates one GPU to the job (lines 13-14). Then, the algorithm computes
the new marginal return of the job and inserts the job back into the queue (lines
15-16). The iterations finish until all GPUs are allocated, or the total power of the cluster exceeds $power\_limit$ (line 10 and lines 18-20).

\parabf{GPU frequency configuration. }
After allocating the GPUs, if the total power of the cluster does not exceed $power\_limit$, \sysname{} will increase the GPU frequencies to further accelerate some jobs.
Similar to GPU allocation, \sysname{} defines another \textit{marginal return} for frequency configuration:
\begin{align} 
	\footnotesize
	priority_F = \frac{(JCT_{j,n,f} - JCT_{j,n,f + \Delta_f})/JCT}{(E_{j,n,f + \Delta_f} - E_{j,n,f})/E}.
\end{align} 

Because the GPU frequency can only be configured as a value from a set of supported values, we increase the GPU frequency by
 $\Delta_f$ each time. The $\Delta_f$ is determined by the hardware. \sysname{} keeps increasing the $f$ of the job with the highest marginal return until all jobs use the highest supported frequency or the total power of the cluster exceeds $power\_limit$ (lines 25-30).
 
\sysname{} re-allocates each job and adjusts the configurations at each \textit{scheduling event}, including job submission, job scaling, and job completion. For each scheduling decision (changing the configuration of a job), \sysname{} checks if each job has been run on this number of GPUs before. If not, \sysname{} profiles all possible GPU frequencies during the following four minutes by partitioning the four-minute time slot into slices at iteration boundaries and dynamically changing the GPU frequency for each slice.  
Meanwhile, \sysname{} further fits the performance models so that the models can help make more accurate performance predictions in the future. 
After the profiling, \sysname{} would receive a job scaling event and adjust the scheduling decisions with the updated performance models.

\subsection{Job Placement}
The job placement mechanism of some existing schedulers~\cite{VavilapalliMDAKEGLSSSCORRB13, XiaoBRSKHPPZZYZ18} might lead to cluster fragmentation, i.e., there are unutilized GPUs in more than one node. We follow previous work~\cite{HwangSharing21} to adopt ``network packing'' and restrict the number of workers of each job to a power of two. It is proved that in this way, at most one job on any node uses two or more nodes, thus avoiding the training performance to be affected by placement~\cite{HwangSharing21}.

Then, we apply buddy allocation~\cite{ZhaoHYZYZYLWXW20} combined with job migration to
eliminate resource fragmentation. The unused nodes are shut down by \sysname{} to avoid extra energy consumption. With this job placement mechanism, we can reduce the energy consumed by the idle GPUs on the nodes that are turned on. 
	\section{Evaluation}
\subsection{Methodology}
\begin{table}[]
	\small
	\begin{tabular}{llll}
		\hline
		Task                                                                            & Dataset                                                          & Model                                                   & Batch Size   \\ \hline
		\multirow{3}{*}{CV}  & \multirow{3}{*}{ImageNet~\cite{imagenet_cvpr09}}                                        & ResNet18~\cite{he2016deep}                                                & 32 - 512 \\ \cline{3-4}
		&                                                                  & VGG16~\cite{SimonyanZ14a}                                                   & 32 - 512 \\ \cline{3-4}
		&                                                                  & \begin{tabular}[c]{@{}l@{}}Inception\\ V3~\cite{SzegedyVISW16}\end{tabular}                                            & 16 - 512      \\ \hline
		\begin{tabular}[c]{@{}l@{}}NLP\end{tabular}                    & aclImdb V1~\cite{ACL-HLT2011}  & \begin{tabular}[c]{@{}l@{}}GPT-2~\cite{radford2019language} \end{tabular} & 8 - 128       \\ \hline
		\begin{tabular}[c]{@{}l@{}}Speech\\ Recognition\end{tabular}                    & LibriSpeech~\cite{PanayotovCPK15} & \begin{tabular}[c]{@{}l@{}}Deep\\ Speech 2~\cite{AmodeiABCCCCCCD16}\end{tabular} & 8 - 256       \\ \hline
	\end{tabular}
	\caption{DNN models used in the evaluation.}
	\label{tab:setting}
\end{table}
\parabf{Simulator.}
To evaluate our scheduler at large scales, we develop a simulator using the profiled information in real NVIDIA V100 GPUs. The simulator simulates all job-level events, including job arrival, scaling,
and completion. We profile the throughputs and energy consumption of each job on real GPU servers as the input of the simulator. 
To make the simulator more realistic, we have also included the pre-run overhead (including JCT overhead and energy consumption overhead) and incorporated it into the simulator. 
The simulator assigns this overhead to each job on job submission.
% Our simulator has very high fidelity, with an error
%rate of no more than 3\% compared with the results in our real cluster experiments.
\begin{figure}[t]
	\centering
	\includegraphics[width=0.9\linewidth]{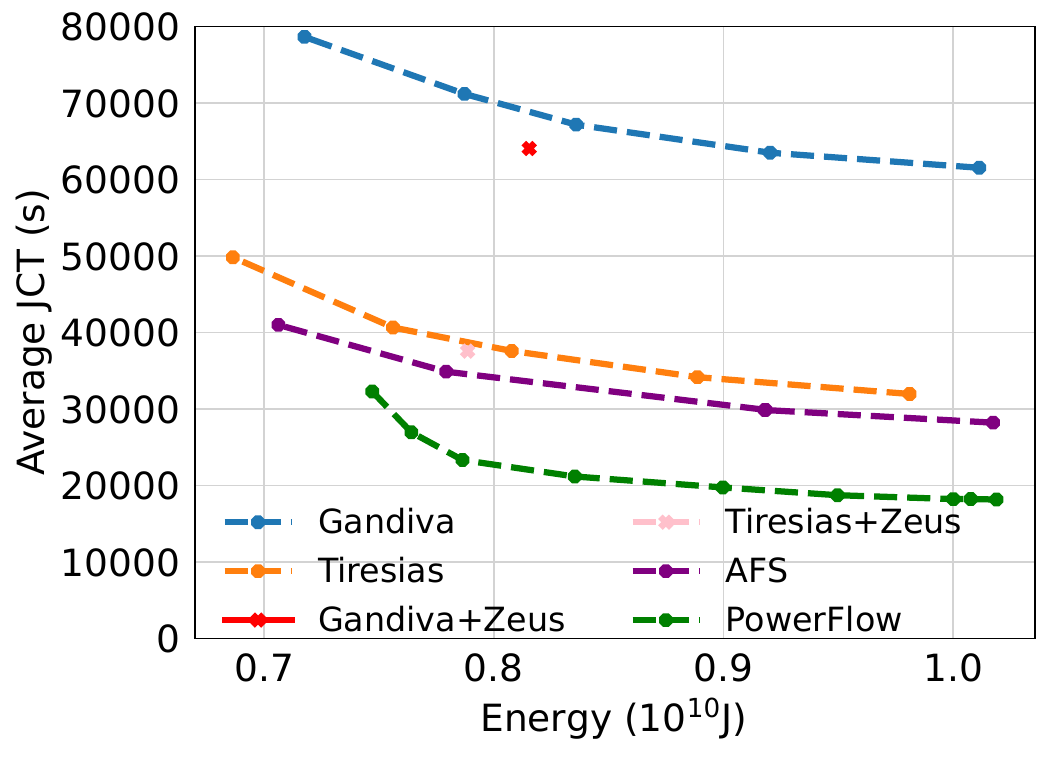}
	\caption{End-to-end scheduling result.}
	\label{fig:e2e}
	%\vspace{-0.1in}
\end{figure}

\parabf{Workloads.}
We use a subset of the public Alibaba trace~\cite{WengXYWWHLZLD22} for end-to-end evaluation. The trace contains 1901 jobs submitted in a 24-hour period, which is substantial compared to existing work~\cite{QiaoCSNH0GX21, GuCSZJQLG19}. Each job in the original trace has information on its submission time, number of GPUs, and duration. However, no information is provided on the DNN architectures being trained, the dataset, and the batch size. 
For each job, we randomly choose a DNN model with a batch size from a pool of representative settings listed in Table \ref{tab:setting}.
%Among the models, the performance of ResNet18 is bottlenecked on Storage IO, the performance of VGG16 is bottlenecked on Network IO, te other models are GPU-dominant. 
Similar to previous
work~\cite{HwangSharing21, NarayananSKPZ20}, we use the duration in the trace
and the pre-measured throughputs to calculate the number of iterations needed to finish each job. 
\begin{figure}
	\centering
	\subfloat[GPU allocation of different schedulers over time.]{
		\begin{minipage}[b]{1\linewidth}
			\includegraphics[width=1\linewidth]{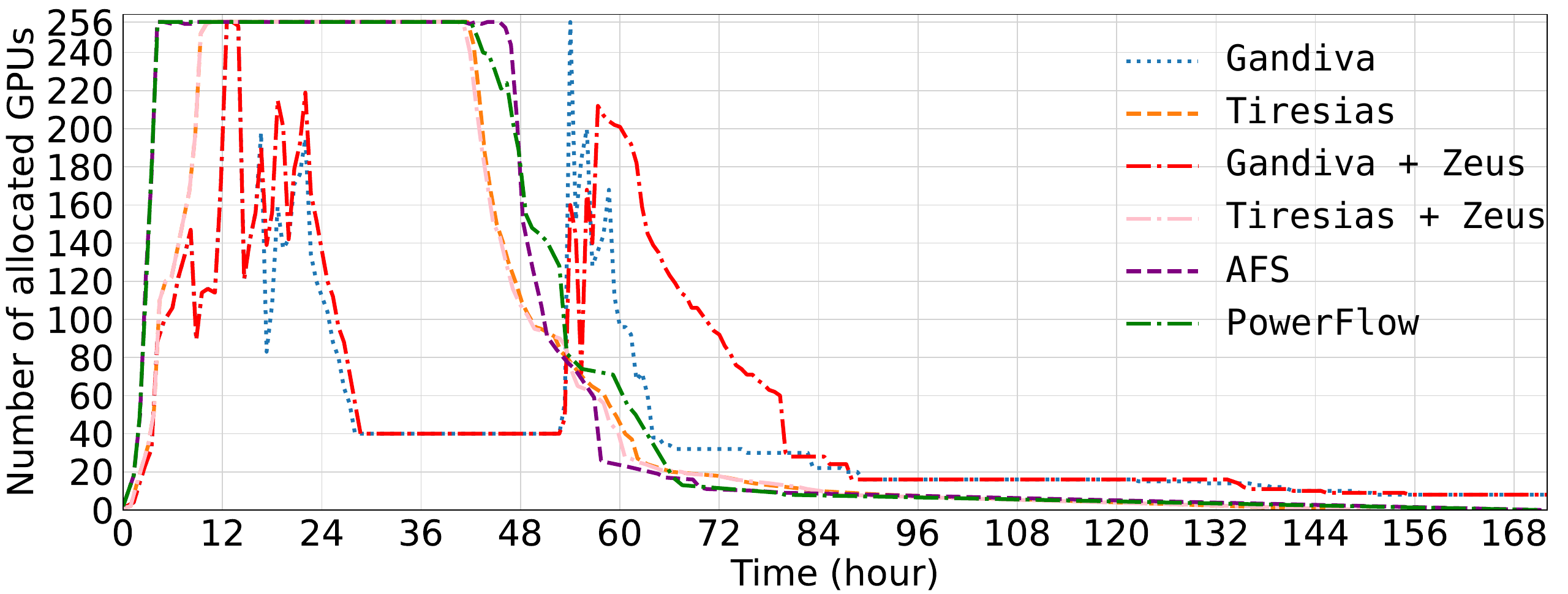}
		\end{minipage}
	}
	\\
	\subfloat[Power of different schedulers over time.]{
		\begin{minipage}[b]{1\linewidth}
			\includegraphics[width=1\linewidth]{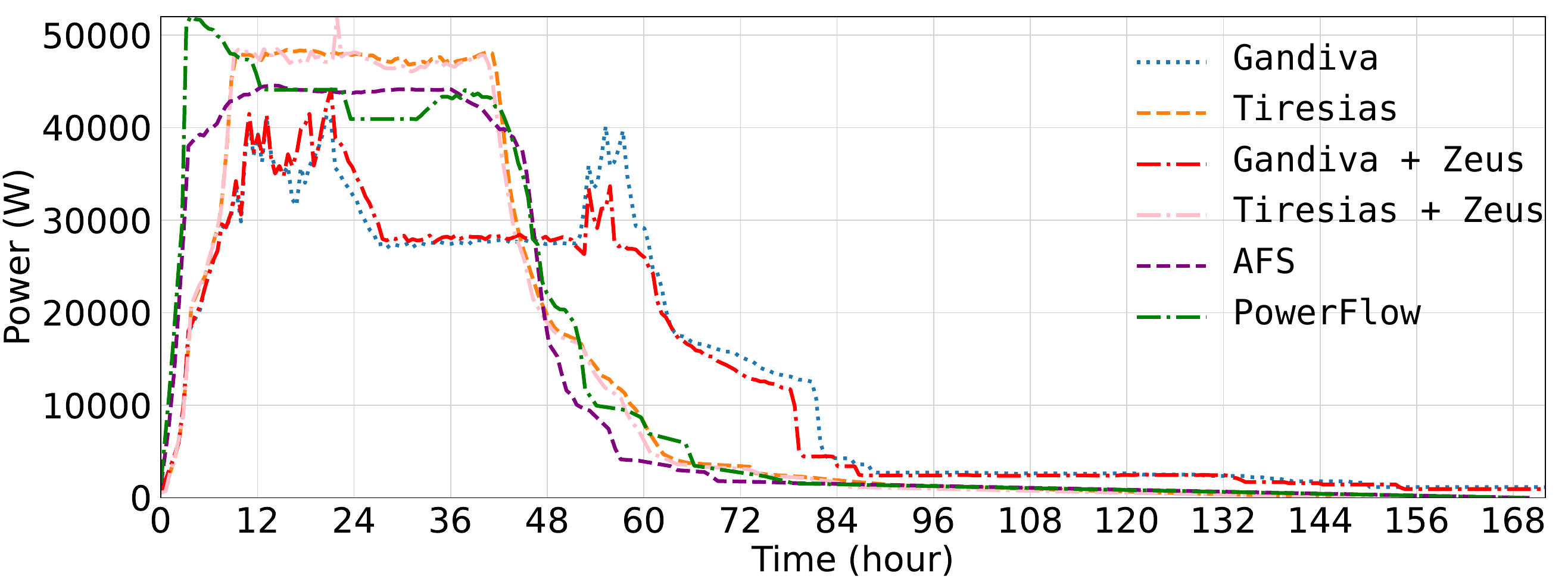}
		\end{minipage}
	}
	%\vspace{-0.1in}
	\caption{Comparison between different schedulers in end-to-end experiments.}
	%\vspace{-0.2in}
	\label{fig:allocation}
\end{figure}

\parabf{Baselines.}We compare \sysname{} to five baselines.
\begin{itemize}%[leftmargin=*]
	\item \textbf{Gandiva:} Gandiva~\cite{XiaoBRSKHPPZZYZ18} is a DL scheduler
	that uses introspective scheduling to refine scheduling decisions
	continuously. It is not elastic (i.e., uses the number of GPUs specified in job traces) and is not energy-aware.
	
	\item \textbf{Tiresias:} Tiresias~\cite{GuCSZJQLG19} uses two-dimensional scheduling algorithms customized for DL jobs. It is also not elastic and is not energy-aware.
	
	\item \textbf{Gandiva + Zeus:} Zeus~\cite{abs-2208-06102} is an optimization framework that finds the tradeoff between energy consumption and training time for a \textit{non-elastic} DL training job. We combine Zeus to Gandiva~\cite{XiaoBRSKHPPZZYZ18} to evaluate this tradeoff in the scheduling of a GPU cluster. 
	It is energy-aware but not elastic.

    \item \textbf{Tiresias + Zeus:} Similar to Gandiva + Zeus,  we combine Zeus to Tiresias~\cite{GuCSZJQLG19}. 
    It is energy-aware but not elastic.	
    
    \item \textbf{AFS:} AFS~\cite{HwangKKSP21} is a scheduling algorithm that accelerates DL training jobs by balancing resource efficiency and short job prioritization. It is elastic but not energy-aware.
\end{itemize}

\parabf{Evaluation metric.} The design goal of \sysname{} is to reduce the average JCT under a certain energy budget. 
Therefore, we compare the average JCT with different energy consumption for all of the mentioned scheduling algorithms.
%Therefore, we compare the average JCT under equivalent total energy consumption of executing all of the jobs in a workload trace. 
We also compare the GPU utilization and the power of the cluster over time.

\subsection{End-to-End Results}
\label{sec:exp:e2e}
Figure~\ref{fig:e2e} compares the average JCT under different energy consumption.
For the baselines that are not energy-aware, we change the total energy consumption of executing all of the jobs by changing the frequencies of all GPU devices. Zeus searches for the energy-efficient configuration for each DL job, so for Gandiva + Zeus or Tiresias + Zeus, we can only get a fixed result with the configurations that Zeus found energy-efficient. As is shown in the figure, \sysname{} achieves the shortest average JCT under different energy consumption. If the GPU frequency is set smaller than 1020MHz for Gandiva, Tiresias, and AFS, they achieve longer JCT while consuming even more energy. Therefore, we do not show these results in the figure.

\begin{table}[]
	\centering
	\begin{tabular}{crr}
		\hline
		DNN Model          & \multicolumn{1}{l}{Throughput} & \multicolumn{1}{l}{Energy} \\ \hline
		ResNet18      & 0.061                     & 0.069                              \\ \hline
		VGG16         & 0.038                      & 0.060                              \\ \hline
		Inception V3  & 0.026                     & 0.045                              \\ \hline
		GPT2          & 0.021                      & 0.068                             \\ \hline
		Deep Speech 2 & 0.045                      & 0.035                              \\ \hline
	\end{tabular}
	\caption{MAPE of the fitted throughput model and energy consumption model.}
	%\vspace{-0.1in}
	\label{tab:model}
\end{table}
\begin{figure}[t]
	\centering
	\includegraphics[width=0.9\linewidth]{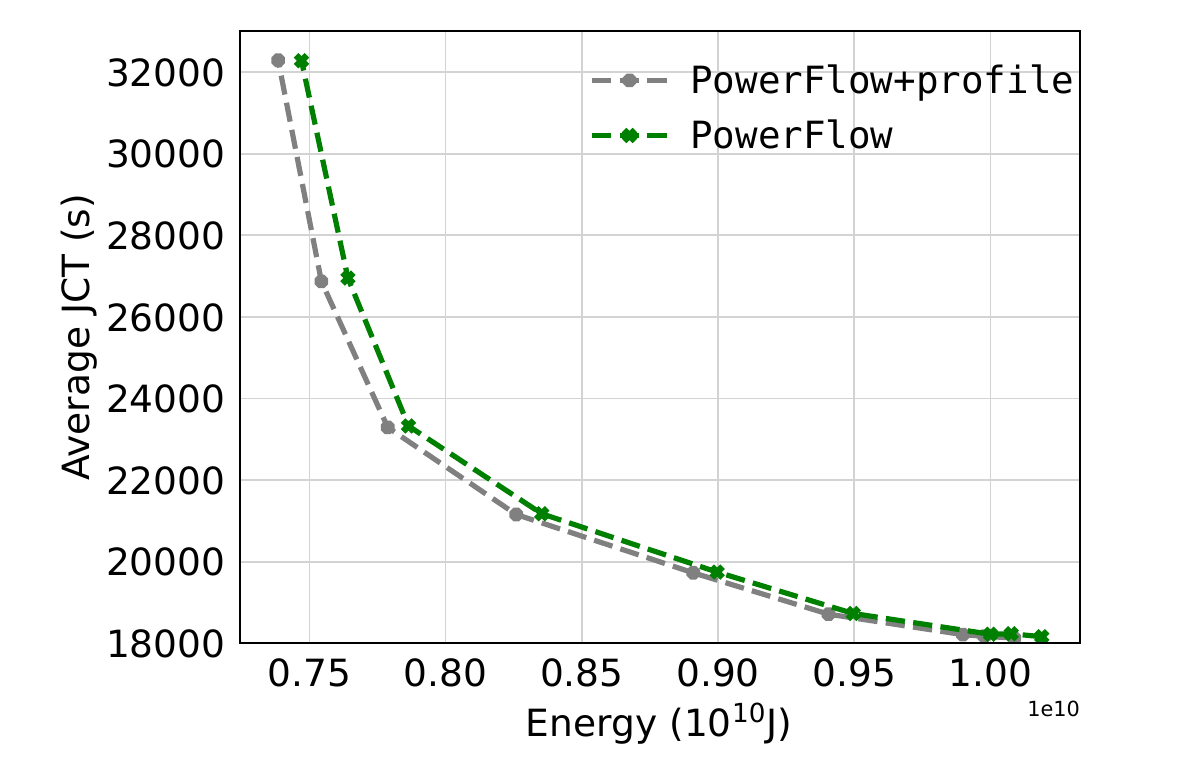}
	\caption{\sysname's scheduling results with pre-profiled performance data and with the performance models.}
	\label{fig:nopred}
	%\vspace{-0.1in}
\end{figure}
\begin{figure*}
	\centering
	\subfloat[Sensitivity to scheduling interval.]{
		\begin{minipage}[b]{0.3\linewidth}
			\includegraphics[width=1\linewidth]{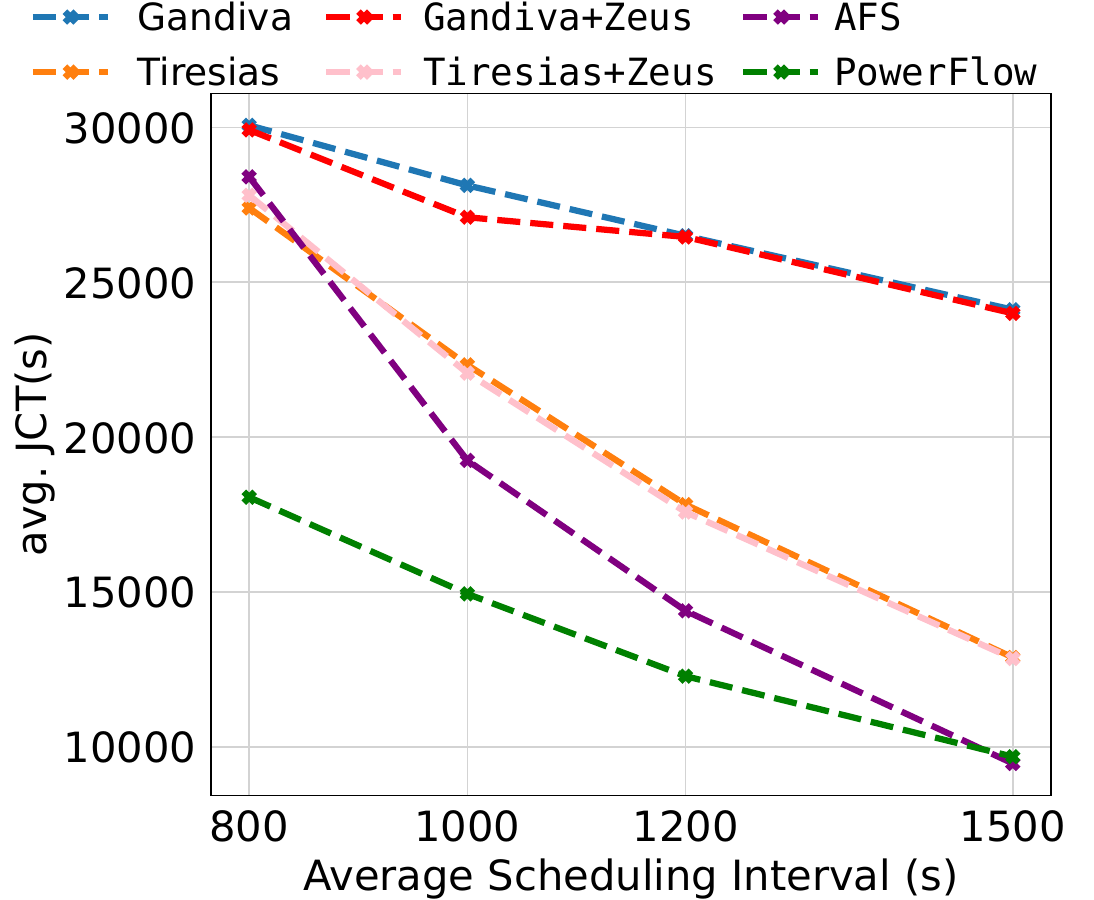}
		\end{minipage}
		\label{fig:sensitivity:a}
	}
	\subfloat[Sensitivity to cluster size.]{
		\begin{minipage}[b]{0.3\linewidth}
			\includegraphics[width=1\linewidth]{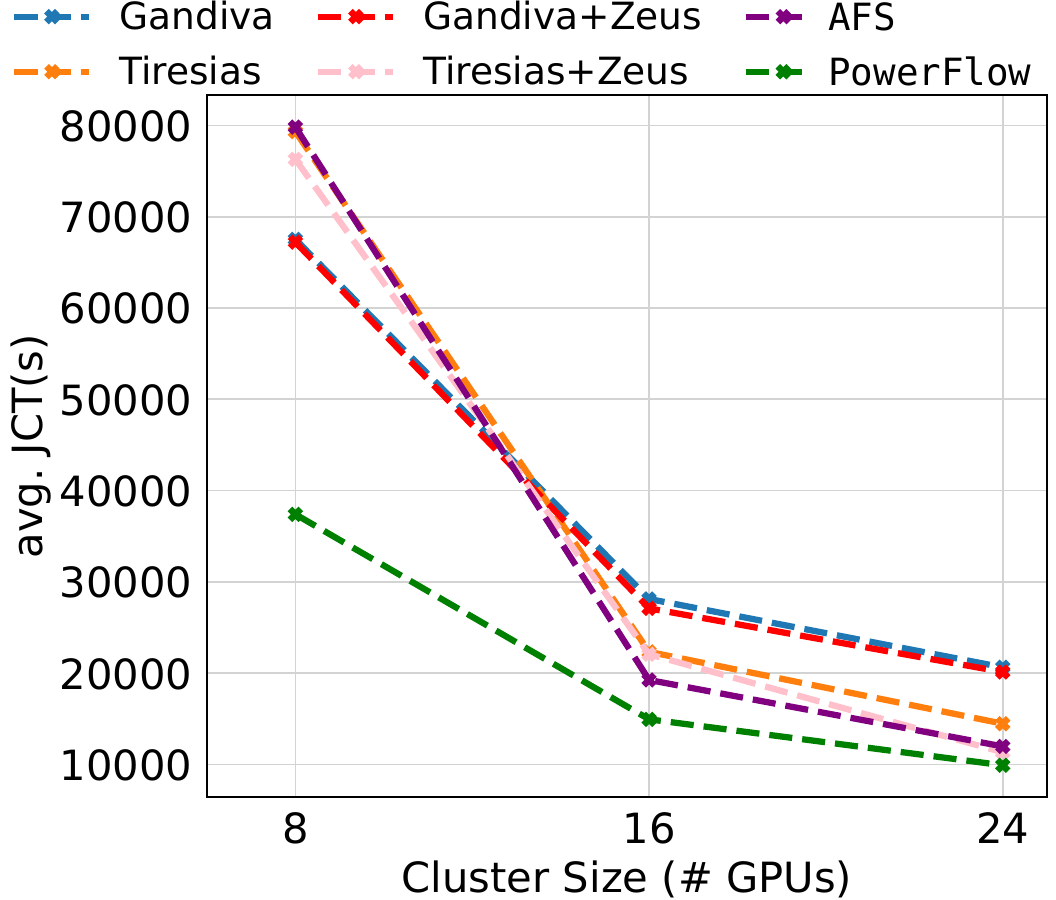}
		\end{minipage}
		\label{fig:sensitivity:b}
	}
	\subfloat[Sensitivity to job size.]{
		\begin{minipage}[b]{0.315\linewidth}
			\includegraphics[width=1\linewidth]{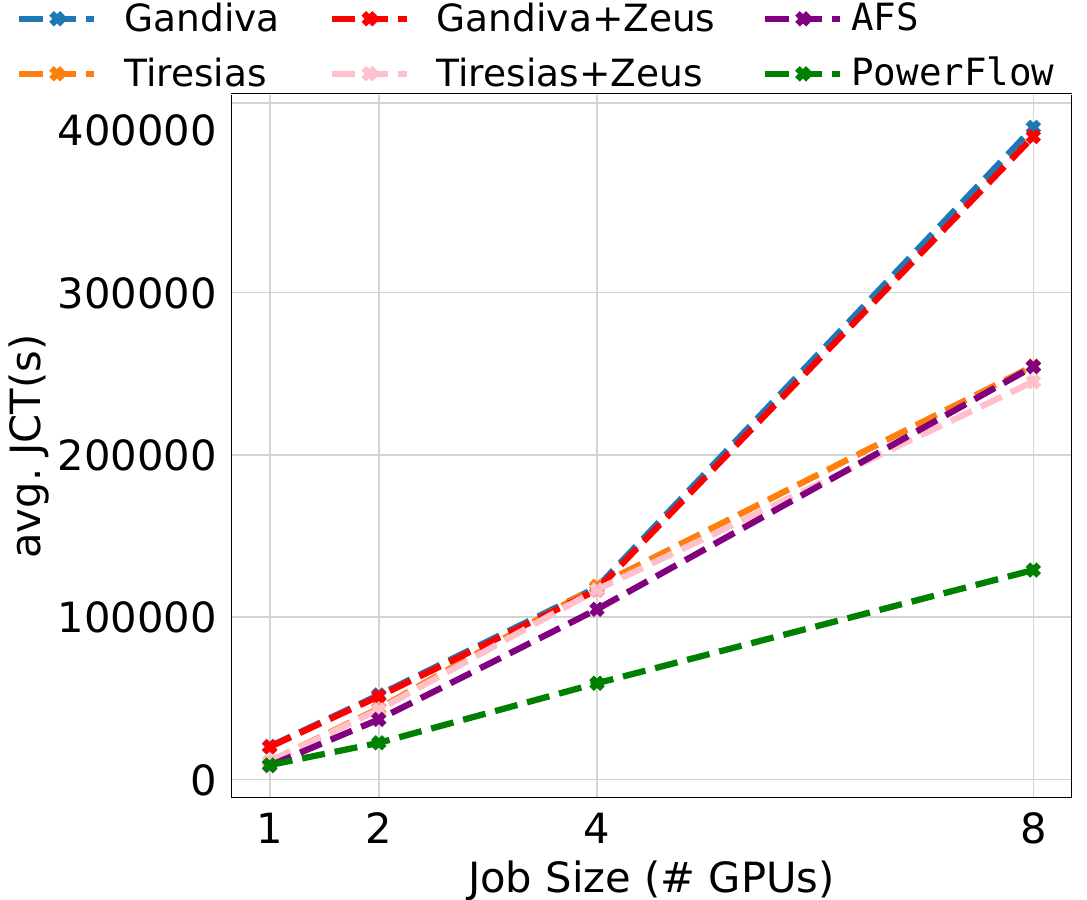}
		\end{minipage}
		\label{fig:sensitivity:c}
	}
	\caption{\sysname's sensitivity to different factors.}
	%\vspace{-0.05in}
	\label{fig:sensitivity}
\end{figure*}

 Compared to Gandiva, Tiresias, Gandiva + Zeus, Tiresias + Zeus, and AFS, \sysname{} improves the average JCT by 3.39$\times$, 1.73$\times$, 2.90$\times$, 1.62$\times$, and 1.57$\times$ at most, respectively.
Gandiva and Tiresias are non-elastic schedulers for DL jobs. 
Compared to them, \sysname{} utilizes more GPUs to accelerate the jobs when there are spare GPUs in the cluster and uses fewer GPUs when the cluster is not large enough to serve all of the submitted jobs. Also, with the placement strategies of Gandiva and Tiresias, there might be fragmentations in the cluster. The energy consumed by these fragmentations also makes Gandiva and Tiresias non-energy-efficient.
Zeus only applies to non-elastic jobs and is not flexible to achieve a shorter JCT if the cloud provider sets a higher energy budget goal.
AFS achieves a rather short JCT because of its elasticity, but it is not energy-aware and can not consume energy more efficiently.

Next, to better analyze the features of different scheduling algorithms, we picked one data point from each of the curves in Figure~\ref{fig:e2e}. Then, we compare the number of GPUs allocated and the power of the cluster.
For the schedulers that are not combined with Zeus, we choose the data point whose total energy consumption is closest to Gandiva + Zeus.
Figure~\ref{fig:allocation} shows the number of allocated GPUs and the power for the chosen data points over time.
From the figure, we observe that  \sysname{} takes full advantage of the idle GPU resources while controlling the total power of the cluster. In this way, the submitted jobs can complete earlier. As is described in Section~\ref{sec:design:algo}, \sysname{} profiles the performance of DL jobs and dynamically adjusts the jobs' resource allocation while fitting the performance models. Therefore, compared to the other baselines, \sysname{} consumes more energy in the first few hours.
Gandiva, Tiresias,  Gandiva + Zeus, and Tiresias + Zeus are not elastic, so they do not utilize idle GPUs if any. 
%When there is a burst of job submissions (e.g., the 13\textsuperscript{th} hour), some jobs are dropped to guarantee the deadlines of admitted jobs.

\subsection{Effects of Performance Models}
\label{sec:exp:model}
Here we evaluate if our throughput model and energy consumption model are accurate enough for \sysname{} to make scheduling decisions. First, we evaluate the Mean Absolute Percentage Error (MAPE) of the two models by using 90\% of the profiled performance values to fit the models and the 10\% left to evaluate the models.
Table~\ref{tab:model} shows the MAPE results. Across the results of all DNN models, the average MAPE of the fitted models does not exceed 10\%, indicating that our models represent the observed performance measurements very well.

Figure~\ref{fig:nopred} compares the results of using our fitted performance models for scheduling and using pre-profiled training performance for scheduling. Note that when scheduling with profiled performance, we assume that the scheduler already knows the performance of each job, so the energy for profiling the performance is not included, which is the ideal case.
We use the same workload in Section~\ref{sec:exp:e2e}  and compare the average JCT as well as total energy consumption. Under the same energy consumption, the difference in average JCT between using profiled performance data and using the performance models is less than 2\%. This indicates that our performance models can be well-fitted for the scheduler to make scheduling decisions. Although using the pre-profiled performance results achieves a slightly shorter average JCT, if we pre-run the performance of all of the configurations offline, it requires too many GPU resources and has an overhead of more than one hour. In reality, this JCT overhead is too large for many DL training jobs, not to mention that the pre-running consumes a lot of extra energy.

\subsection{Sensitivity Analysis}
In this section, we evaluate \sysname's sensitivity to different factors. Following previous work~\cite{NarayananSKPZ20}, we generated 100-job traces randomly for sensitivity analysis.
As we explained in Section~\ref{sec:exp:e2e}, unlike other schedulers that have different scheduling results when the GPU frequency is set to different values,
Zeus only chooses one energy-efficient configuration for each job. 
We observe that when scheduling jobs from the same trace in the same cluster, Gandiva + Zeus and Tiresias + Zeus achieves similar total energy consumption (less than 5\% difference).
To make a fair comparison, for the schedulers that are not combined with Zeus,
we choose the scheduling results that have comparable total energy consumption with Gandiva + Zeus and Tiresias + Zeus (less than 5\% difference). 
Then, we compare the average JCT achieved by these schedulers.  

\parabf{Sensitivity to scheduling interval.}
First, we compare the average JCT of \sysname{} and the baselines under different loads in terms of average job arrival interval. Figure \subref*{fig:sensitivity:a} shows the results.
As expected, for all schedulers, the average JCT gets shorter as the interval increases.
Compared with the baselines, \sysname{} always achieves a short average JCT and has a larger JCT improvement with a heavier cluster load. This is because \sysname{} allocates the limited GPUs to the jobs that can bring higher energy efficiency to the cluster.

\parabf{Sensitivity to cluster size.}
Given the same set of jobs, the average JCT is shorter in a larger cluster with more GPUs. We run \sysname{} and the
baselines on the same job trace with different cluster sizes.
Figure~\subref*{fig:sensitivity:b} shows that \sysname{} outperforms the baselines
more in small clusters. This is because in a large cluster, both \sysname{} and AFS can scale out the jobs to utilize more GPUs for average JCT improvement; but in a small cluster, \sysname{} utilizes the GPUs more efficiently.

\parabf{Sensitivity to job size.}
In today's DL training platforms, users usually specifically allocate a fixed number of GPUs to the jobs. Both Gandiva and Tiresias require DL developers to specify the number of GPUs for DL jobs.
We evaluate the average JCT on the same cluster size when job size varies.
As expected, from Figure \subref*{fig:sensitivity:c}, we can see that for all scheduling algorithms, the average JCT is smaller with small jobs.
For larger jobs, we observed a larger improvement by \sysname{} compared to the baselines. 
This indicates that training some DNN models with a small number of GPUs is more energy efficient than training
with a large number of GPUs and default GPU efficiency.

	\section{Related Work}
\parabf{Scheduler for DL jobs.}
Early efforts used cluster managers like
Kubernetes or YARN to schedule DL jobs in the cloud without considering the
characteristics of DL jobs, which results in low
performance~\cite{boag2017scalable, jeon2018multi, JeonVPQXY19}. Recent efforts
proposed specialized cluster schedulers for DL training jobs~\cite{XiaoBRSKHPPZZYZ18, GuCSZJQLG19, peng2018optimus, gandivafair, themis, QiaoCSNH0GX21}.
These efforts focus on optimizing for JCT, fairness, or GPU utilization while ignoring the energy consumption of DL jobs and the GPU cluster.

\parabf{Energy measurement for DL jobs.}
Existing studies have investigated  the measurement or estimation of the carbon emission and energy consumption of machine learning or DL jobs~\cite{abs-2104-10350, abs-2002-05651}.
Similar to these studies, \sysname{} measures the energy consumption on GPU via NVML~\cite{nvml} and then estimates the energy consumption with different configurations with \sysname's performance models.

\parabf{Energy optimization for DL jobs.}
Several efforts have studied the impact of GPU DVFS and power configuration on the performance of GPU devices and DL jobs~\cite{Mei0017, TangWWC19}. These methods require offline profiling or modeling, which is not realistic or brings huge overheads to online cluster schedulers.
Zeus~\cite{abs-2208-06102} is an online optimization framework for recurring DL training jobs that finds the trade-off between throughput optimization and energy consumption by automatically tuning the batch size and GPU power limit of training jobs. However, it does not apply to elastic jobs and lacks the view of the whole cluster if applied to a GPU cluster scheduler directly.

\parabf{Energy efficient VM scheduling.}
Energy consumption optimization in traditional datacenters has been studied by a recent line of research work. Some studies leverage DVFS based on system performance requirements at the given time~\cite{WangLL21}. This requires fine-grained resource sharing, which is undeveloped and unstable in today's GPU usage. There are also 
energy consumption optimization schedulers based on workload forecasting~\cite{wang2018robust,RanjbariT18, tang2018energy}. These methods either require user-defined SLA, which does not apply to today's common practice of DL training, or balance workloads and shut down unused servers, which is similar to \sysname{}'s job placement mechanism.
	\section{Conclusion}
In this paper, we proposed \sysname{}, an energy-aware scheduler for GPU clusters that reduces the average
JCT under an energy budget with the tradeoff between the average JCT and total GPU energy consumption.
\sysname{} predicted the throughput and energy consumption performance of DL training jobs with performance models.
We developed a scheduling algorithm that dynamically allocates GPUs to submitted jobs and adjusts the GPU frequencies based on the performance predictions made by the performance models.
\sysname{} applied network packing and buddy allocation to job placement, avoiding extra energy consumption of cluster fragmentations.
The evaluation results showed that under the same energy consumption, \sysname{} improved the average JCT by 1.57 - 3.39$\times$ at most, compared to competitive baselines. 
	\label{lastpage}
	
	\bibliographystyle{plain}
	\bibliography{reference}
	\label{allpages}
	
	%%%%%%%%%%%%%%%%%%%%%%%%%%%%%%%%%%%%%%%%%%%%%%%%%%%%%%%%%%%%%%%%%%%%%%%%%%%%%%%%
\end{document}